# A front-end electronic system for large arrays of bolometers

**C.Arnaboldi[a, b], P.Carniti[a, b] L.Cassina[a, b], C.Gotti[a, b], X.Liu[c], M.Maino[a, b], G.Pessina[a,b], C.Rosenfeld[c], B.X.Zhu[d]**

[a] *INFN – Istituto Nazionale di Fisica Nucleare, sezione di Milano-Bicocca, Piazza della Scienza 3, Milano, 20126, Italy*

[b] *Università di Milano-Bicocca, Dipartimento di Fisica G. Occhialini, Piazza della Scienza 3, Milano, 20126, Italy*

[c] *Department of Physics and Astronomy, University of South Carolina, Columbia, SC 29208 USA*

[d] *Department of Physics and Astronomy, University of California, Los Angeles, 475 Portola Plaza, Los Angeles, CA 90095, USA*

*E-mail*: Paolo.Carniti@mib.infn.it

ABSTRACT: CUORE is an array of thermal calorimeters composed of 988 crystals held at about 10 mK, whose absorbed energy is read out with semiconductor thermistors. The composition of the crystal is $TeO_2$, and the aim is the study of the double beta decay of $^{130}$Te on very long and stable runs. CUPID-0 is an array of 26 Zn$^{82}$Se crystals with double thermistor readout to study the double beta decay of $^{82}$Se. In the present paper, we present an overview of the entire front-end electronic readout chain, from the preamplifier to the anti-aliasing filter. This overview includes motivations, design strategies, circuit implementation and performance results of the electronic system, including other auxiliary yet important elements like power supplies and the slow control communication system. The stringent requirements of stability on the very long experimental runs that are foreseen during CUORE and CUPID-0 operation, are achieved thanks to novel solutions of the front-end preamplifier and of the detector bias circuit setup.

## Contents

---

## 1. Introduction and functional block diagram

The study of rare radioactive decays using materials that serve both as source and detector was introduced in the 1960's [1], [2]. Such studies are generally conducted at sites deep underground in order to suppress cosmogenic background [3]. The ideal detector for such experiments is a pure calorimeter, and cryogenic detectors are well adapted to this role [4], [5], [6], [7], [8]. CUORE (Cryogenic Underground Observatory for Rare Events) [9] is an array of 988 $TeO_2$ crystals held at a temperature around 10 mK. The primary goal of CUORE is the observation of neutrinoless double beta decay of the isotope $^{130}Te$, which is present in the crystals with its natural abundance of 34%. Another experiment, which takes a similar approach using 26 $Zn^{82}Se$ crystals but adds the possibility to identify particles, is CUPID-0 (CUORE Upgrade with Particle Identification and previously called LUCIFER) [10]. It is a demonstrator for the next phase of CUORE.

In both experiments, when a particle, whether generated internally or externally, propagates in a crystal, the crystal warms. As a result the impedance of a semiconductor thermistor glued to the crystal changes. If a small DC current biases the thermistor, the change in voltage across it can be amplified and measured. In CUPID-0, the direct calorimetric signal is augmented with a second indirect signal. A portion of the light generated by scintillation along the particle path is absorbed in a thin slab of Ge optically coupled to one face of the $TeO_2$. In consequence a thermal pulse occurs also in the Ge. A second thermistor glued to the Ge slab generates an electrical pulse, which propagates through an independent chain of electronics. The intensity of the scintillation light may differ significantly between signal and background processes. Thus the Ge signal can enhance background suppression. This paper describes in detail the chain of electronics that manipulates these bolometric signals as they travel from the thermistors to the data acquisition system.

We will not discuss further details of the physics of the detectors but only list the main characteristics they impose on the electronics. The signal bandwidth is small, a few Hz, because the heat capacitance of the 0.76 kg crystals is large and the thermal conductance to the heat reservoir is small by design. As a consequence: **1)** the detector and all the amplifying stages are DC coupled (a convenience also for baseline studies). Bolometers have a large spread in both the impedance value and energy conversion gain (factors of two to three are typical, 100 μV/MeV to 300 μV/MeV), and several consequences follow: **2)** the amplifying system must have adjustable gain to exploit the full dynamic range of the data acquisition system (DAQ); **3)** the biasing of the bolometer must be independently adjustable from detector to detector to maximize signal amplitude (which is expected to range from a few thousands up to ten thousand); **4)** noise inserted by the electronics must be adequately low (a few nV/√Hz for series noise and a few fA/√Hz for parallel noise are the levels). The double beta decay is a very rare process and the observation time needs to be very long (time scale is years). Therefore, **5)** the front end must be very stable in gain and offset and **6)** show very small thermal drift (below 1 μV/°C). With adequately small electronic drift long-term changes in the energy conversion gain of a crystal can be detected and compensated by tracking the bolometer baseline voltage in conjunction with a pulser stabilization system.

To convey a simplified view of the CUORE system we describe it with several subsystems as shown in Figure 1.

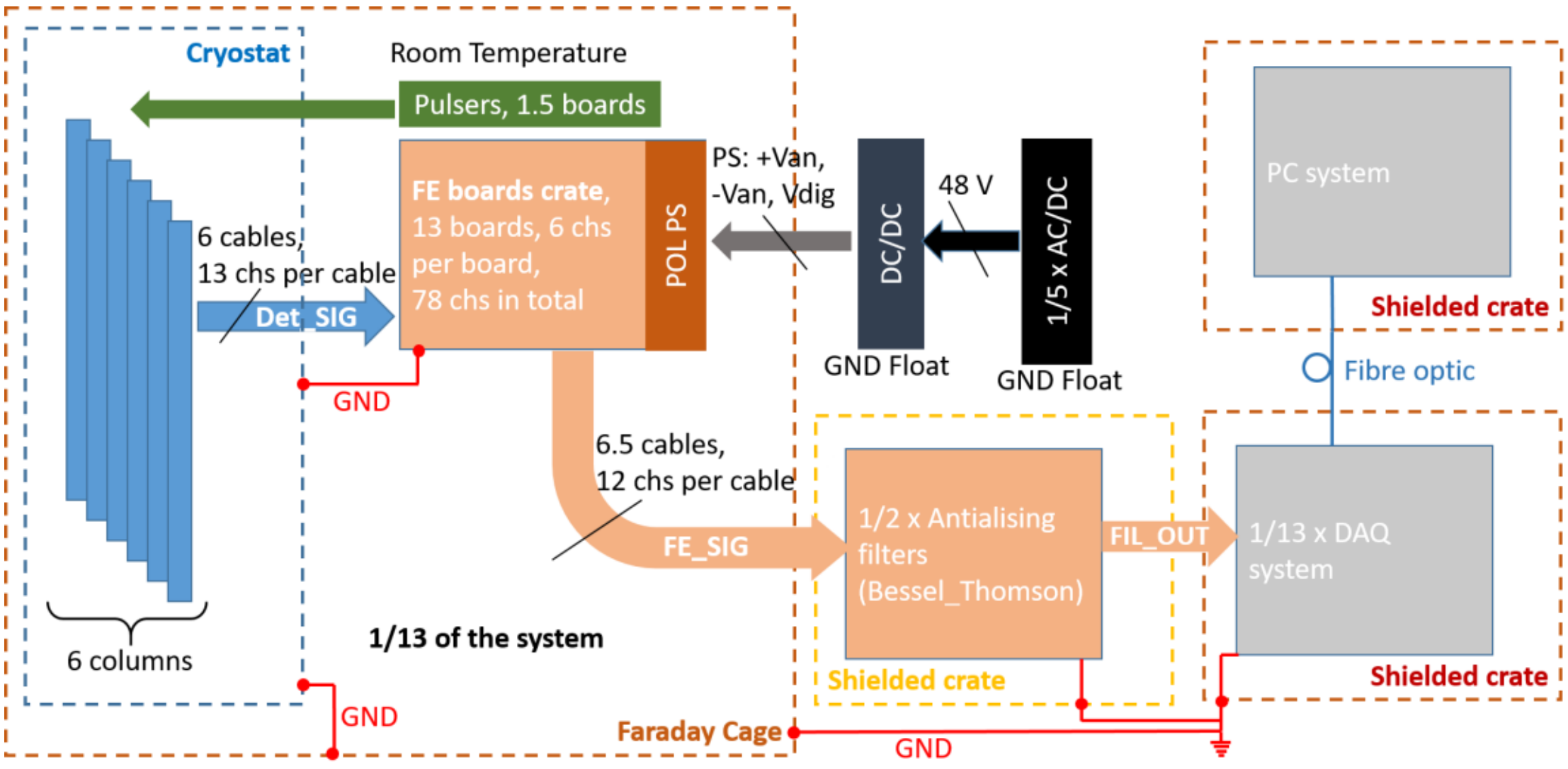

**Figure 1: Simplified scheme of one module out of 13 for CUORE (the quantities indicated are those needed to equip one module). 78 crystals assembled as six columns are read out (arrow path labelled Det_SIG) via the FE boards which are housed in a crate. The outputs from the FE boards (arrow labelled FE_SIG) feed the anti-aliasing filters. The filter outputs (arrow labelled FIL_OUT) in turn feed the DAQ, which implements analog-to-digital conversion. The PC system pulls data from the DAQ through a fibre optic link. Every part of the system is in communication with the control room via fibre optic links (not shown). The ground, GND, has the hub at the DAQ and is common to the anti-aliasing filter crates and the Faraday cage. The cryostat and the FE crates are grounded via the Faraday cage.**

## 2. Geographic arrangement of system components

The layout of the very front-end part of the system, Figure 1, is extremely important being close to the detectors, the source of the signals. It is organized in crates (standard 19 inch - 6 units height) containing each 13 FE boards, or mainboards, each with 6 front-end amplifiers, and one power supply unit, POL PS. The inputs to a crate are wire pairs gathered into six cables each of which serves the 13 thermistors of one column of crystals. Surrounding the 14 crates of FE boards, the pulser crates and the cryostat is a Faraday cage [12] (although 13 crates would be sufficient for the crystals, additional FE boards process auxiliary signals for diagnostic thermometry). A two stages AC/DC floating system [13] supplies the power supply units POL PS. Two shielded crates hosting the anti-aliasing filters, two hosting the DAQ boards, and one for the DAQ PC are situated immediately outside the Faraday cage. To suppress EMI and ground loop fluctuations [14] [15], the DAQ boards communicate with the DAQ PC system via optical fibre. This fibre optic link does not, however, allows the DAQ boards to float isolated from the main ground. This circumstance makes the ground, GND, at the DAQ crates the most suitable site for the hub of a star configuration, to which the anti-aliasing filter crates and the Faraday cage are connected. The cryostat and the FE crates are grounded to the DAQ through the Faraday cage. Circuitry on the FE boards applies a DC biasing current to the thermistors via the connecting wire pair and via the same wire pair accepts the voltage pulses from the detectors, which constitute the signal. The FE boards amplify the pulses and convey them to the anti-aliasing filters whose outputs are then digitized by the DAQ and recorded in the DAQ PC system.

Above each stack of FE crates is another crate, 19 inches wide by three units high, which houses the pulsers. Each channel of the pulser boards [17] [18] can create a thermal pulse in one of the crystals by the application of a precise and stable pulse to a resistor glued to one face of the crystal [15]. These pulses emulate the energy release by particles in the crystals. In conjunction with tracking of the baseline voltages of the crystals, the pulser system enables detection and compensation of the long-term drift in the energy conversion efficiency of individual crystals.

## 3. The front-end crate

### 3.1 Overall crate organization

The front-end crate of Figure 1 has 13 boards, 233×280 $mm^2$, 8 layers. On one side of the crate is the Point-Of-Load Power Supply, POL PS, composed of two independent linear power supplies featuring very low 1/f noise and high stability [19]. The POL PSs operate also as a voltage reference.

The six cables that connect the detectors arrive at the backplane of the FE crate. The first 12 channels of each cable are routed to two adjacent boards, while the last channel is routed to the thirteenth board of the crate. This layout allows a degree of regularity in the connections. On the backplane only the signals from the detectors and the bias currents are present. No voltage supplies or digital signals are present there to minimize electronic interference.

### 3.2 The front-end board organization

Figure 2 is a photograph of the mainboard. On the left are the two input connectors. The impedance of the thermistors is generally in the hundreds of MΩ range and occasionally exceeds a GΩ. Thus to preserve the integrity of the bias currents and signals the parasitic impedances to ground or to the electrodes of adjacent channels should be at a minimum hundreds of GΩ. The electrical link from the thermistors to the input connectors of Figure 2 was extensively studied [20], [21]. It consists of three parts, two of which are inside the cryostat (the first from the detectors to the mixing chamber, the second from the mixing chamber to the top of the cryostat) [22], [23]; while the last is at room temperature and connects the top of the cryostat to the front-end inputs. The overall length is 5 m on average. The resulting parasitic capacitance in parallel with each thermistor is about 500 pF, while the parasitic resistance is indeed in the requisite hundreds of GΩ range. At the operating temperatures of CUORE, around 10 mK, the detector bandwidth is around 1 Hz. Therefore, the parasitic capacitance of the link limits the bandwidth when the detector dynamic impedance is larger than a few hundreds of MΩ

The signals from the thermistors are differential, which further enhances immunity to inter-channel coupling.

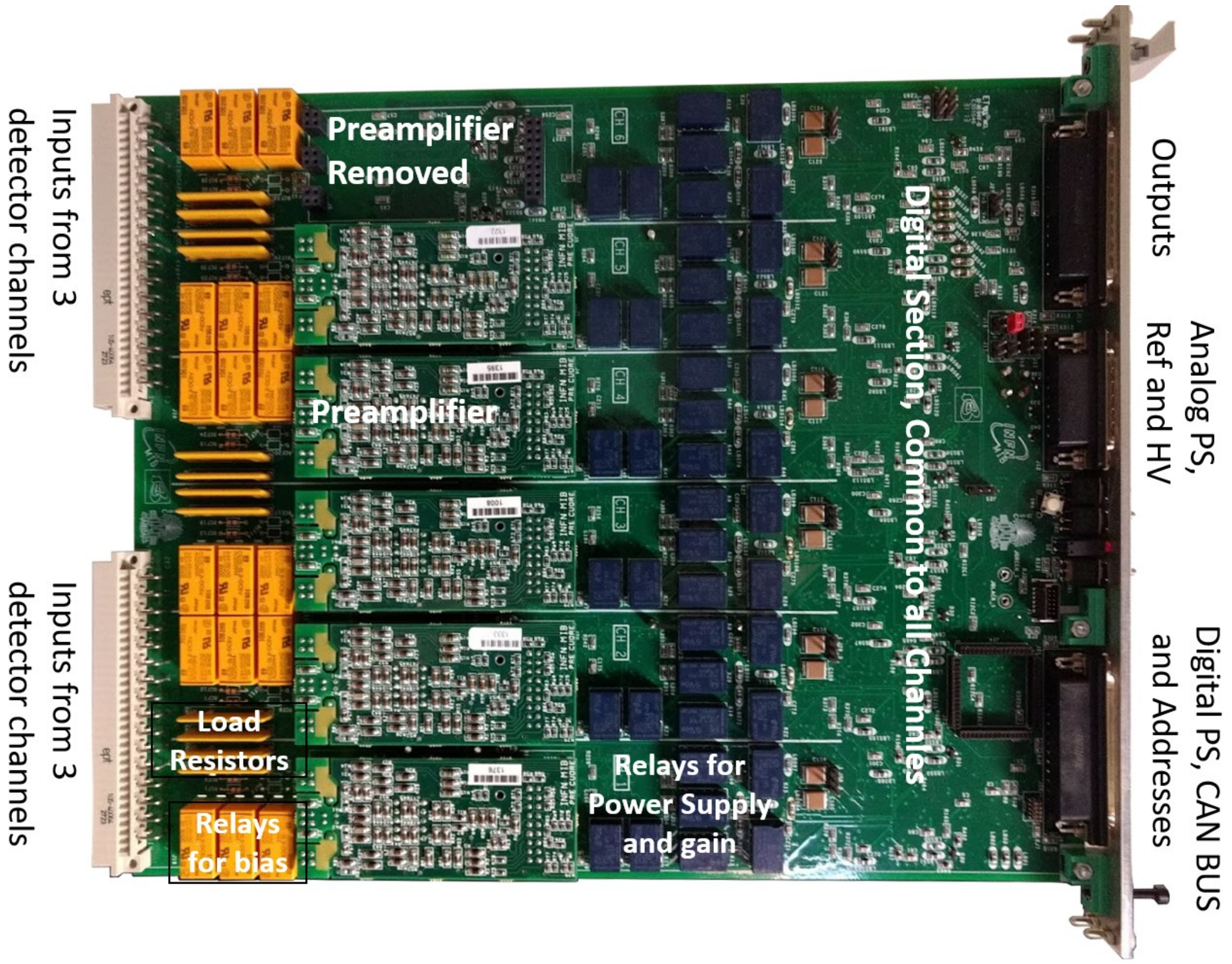

**Figure 2: Photograph of the top of the mainboard. The preamplifier on the top of the figure was removed.**

In Figure 2 we see adjacent to the input connectors load resistors, relays, and other circuits for thermistor biasing. To the right of the biasing circuits are the preamplifiers. Stepping to the right of the preamplifiers we see relays that enable/disable power to each channel and other relays that select the gain, x1 or x10, of the first stage of the Programmable Gain Amplifiers (PGAs). The PGAs are mounted on bottom side of the board beneath the relays and are not visible in the picture. Between the relays and the connectors at the right edge of the board is digital circuitry that serves all of the six channels.

The six differential outputs from the PGAs are routed to the 25-pin female D-Sub connector at the upper right. Power for digital circuitry, communication with the control room, and the board address arrive on the 25-pin male D-Sub connector at lower right. Power for analog circuitry and for thermistor biasing arrive on the 15-pin male D-Sub connector at middle right.

### 3.3 The thermistor biasing system

The energy conversion gain of cryogenic detectors has a large spread (a factor of 2 to 3 is typical) and must be determined *in situ*. It is strongly dependent on the biasing. The first optimization step for a channel is the measurement of the load curve followed by the noise measurement and the pulse response. These measurements yield a determination of the Signal to Noise ratio, S/N, as a function of the applied DC bias. This process is automated.

Figure 3 shows the circuit diagram for detector biasing.

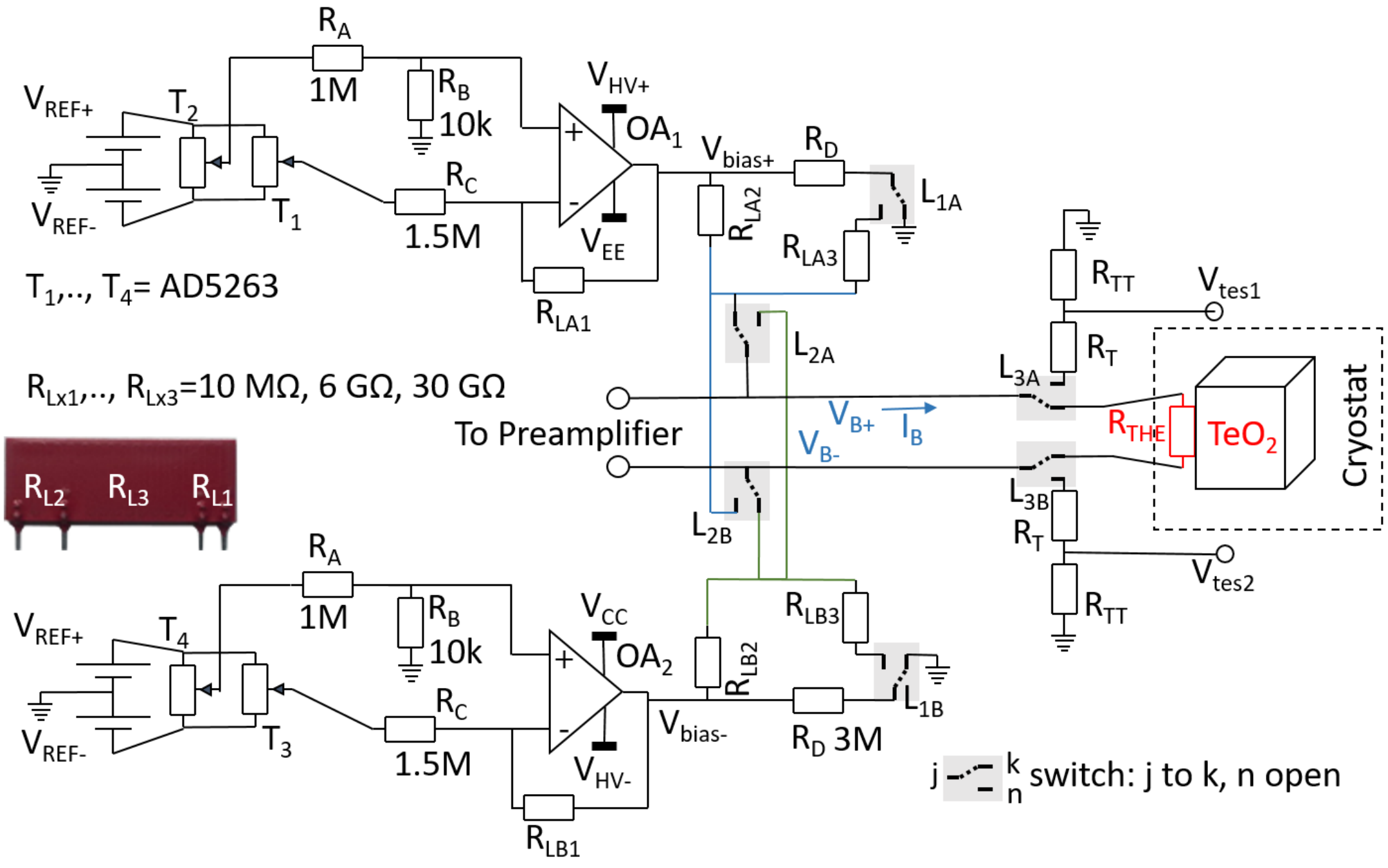

**Figure 3: Schematic of the biasing circuit for the detector thermistor, $R_{THE}$, glued on the $TeO_2$ crystal, not to scale.**

The thermistor is biased in a differential way. We generate the symmetric voltages $V_{bias+}$ and $V_{bias-}$ with the amplifiers $OA_1$ and $OA_2$ (OP140). The values of the voltages are set by the two pairs of programmable trimmers $T_1$ and $T_2$, $T_3$ and $T_4$, all part of the same AD5263. $T_1$ ($T_3$) sets the coarse value, while $T_2$ ($T_4$) tunes the fine value ($T_2$ and $T_4$ are attenuated by 100 V/V with $R_A$ and $R_B$). $V_{bias+}$ is settable from 0 V to +25 V and $V_{bias-}$ from 0 V to -25 V. The asymmetric power applied to $OA_1$ and $OA_2$ ensures that the 36 V absolute maximum between their rails is not exceeded.

The actual bias $V_{B+}$-$V_{B-}$ applied to the detector is a fraction of $V_{bias+}$-$V_{bias-}$ due to the partition coming from the load resistor system, $R_{LA2}$ and $R_{LB2}$. Let us consider, for the moment, the relay switches of Figure 3 connected as indicated. The range of values needed for the load resistors is obtained by analysing their parallel noise contribution, which should satisfy:

$$\frac{4K_B T_R}{R_L} R_{THEDyn}^2 \leq 4K_B T_D R_{THEDyn} \quad \rightarrow \quad R_L \geq \frac{T_R}{T_D} R_{THEDyn} \tag{1}$$

In (1) $K_B$ is the Boltzmann constant, $R_L$= $R_{LA2}$ + $R_{LB2}$, $T_R$ is room temperature, $T_D$ is the temperature of $R_{THE}$, and $R_{THEDyn}$ is the dynamic impedance of $R_{THE}$ [24], [25], [26], [27]. $R_{THEDyn}$ is a fraction of $R_{THE}$. With $T_R$ at 300 K, $T_D$ at 15 mK, and $R_{THEDyn}$ a few tens of MΩ the $R_L$ in (1) would ideally be greater than $10^{12}$ Ω. For practical reasons, stemming from the maximum voltage available for biasing and suppression of parasitic impedance, we chose $R_L$ to be 60 GΩ.

If the amplifier and load resistors were closer to the bolometer temperature we could have adopted a more favourable value for $R_L$. The "cold" solution has been investigated by others [29] [30], and we undertook a preliminary study for CUORE [28]. Ultimately we opted for the room temperature solution. It carries the benefit of easier implementation and maintenance, and even with $T_R$ at room temperature the intrinsic detector noise is not the dominant limitation on the energy resolution. The contribution from mechanical friction is as or more significant [31] [32] [33].

From (1) we have that $R_L$ is much greater than $R_{THE}$, therefore the biasing system behaves like a current generator:

$$I_B = \frac{V_{REF+} - V_{REF-}}{R_C}\frac{R_{LA1}}{R_L}\left\{\alpha_3 - \alpha_1 + (\alpha_2 - \alpha_4)\left(\frac{R_B}{R_A + R_B}\right)\left(\frac{R_C}{R_{LA1}} + 1\right)\right\} \qquad (2)$$

In (2) $\alpha_1$,… $\alpha_4$ are the trimmer setting fractions, 0 ÷ 1, for $T_1$… $T_4$ (half scale corresponds to $I_B$= 0 A); and $R_{LB1}$ and $R_{LB2}$ have been taken equal to $R_{LA1}$ and $R_{LA2}$. If the switches $L_{1A}$ and $L_{1B}$ are toggled to the positions opposite to the ones shown in Figure 3, then $R_L= (R_{LA2}||R_{LA3})+(R_{LB2}||R_{LB3})$ ($R_X||R_Y$ is the resistance of $R_X$ and $R_Y$ in parallel), and (2) remains valid as written. Voltages $V_{REF}$ and trimmers $T_x$ are all stable at the level of a few ppm/°C, while resistors $R_A$, $R_B$ and $R_C$ are metal film (mini-melf), with less than 10 ppm/°C drift. Resistors $R_{L2}$ and $R_{L3}$ have large resistances, 6 GΩ and 30 GΩ, and are not available with the favorable qualities of metal film; nevertheless they enter in (2) as their ratio, so that the important feature is their relative tracking. To optimize the tracking of the three resistors $R_{LX}$ we designed a custom array. A photograph of a sample is shown at middle-left of Figure 3. We selected two companies[1] for fabricating the array. Results from both were extremely good, and we use indifferently samples from each company. We measured the tracking ratio between the resistors to be better than 10 ppm/°C. In addition to thermal drift we must also be concerned about the low frequency (LF) or 1/f noise, which is important when large bias voltages are applied. By making the resistor long, about 20 mm in the case of $R_{L2}$, the electric field per unit length is minimized, which helps to reduce the LF noise [34].

In Figure 3 the setting indicated for the relays is for the usual data-taking condition. A number of other combinations are available and exploited during the characterization of the thermistors. Each relay on the board has a pair of double-pole single-throw switches and is bi-stable (its coils need to be powered only during state transitions). The relay model chosen for $L_1$, $L_2$ and $L_3$ (DS2E-SL2-DC5V) has a relatively large spacing of 5.08 mm (200 mils) between terminals, which helps to maintain a high value of the parasitic resistances. In addition, several small apertures in the board have been placed around all the terminals needing large parasitic impedances as one sees Figure 4. These break possible conductive paths due to solder flux residues and dust.

[1] Ohmcraft and SRT Resistor Technology

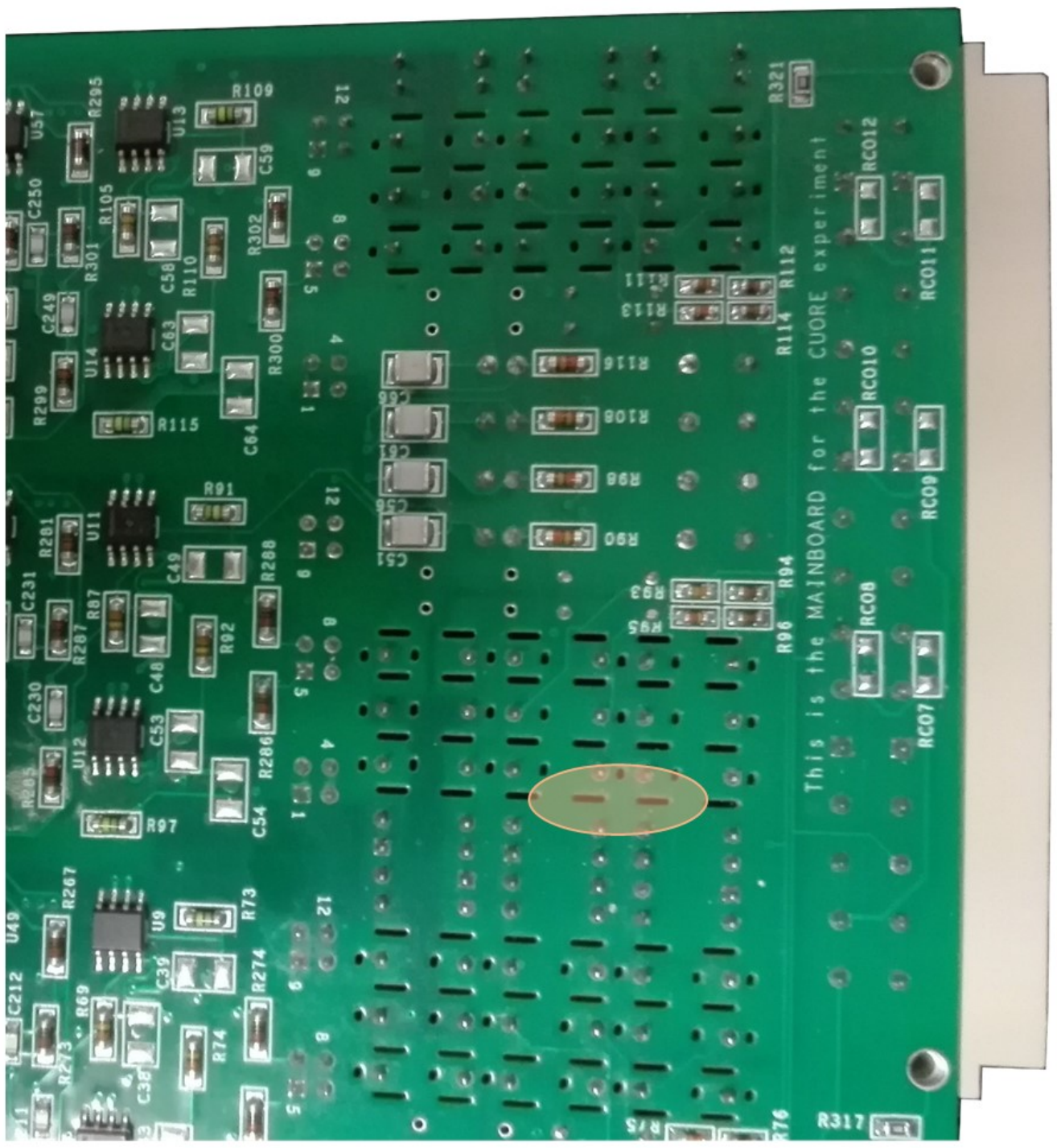

**Figure 4: Zoom of a region of the bottom side of the mainboard around the inputs. The orange oval highlights two of the several apertures placed close to terminals where a high impedance is needed. The apertures break parasitic conductive paths on the surface of the board.**

Switches $L_{2A}$ and $L_{2B}$ are used to reverse the polarity of the bias applied to the thermistor. Bias reversal is useful during the characterization phase because the difference between the measurements taken with opposite bias polarity is independent of any offset. When $L_{3A}$ and $L_{3B}$ are switched to their alternative position, the bias current is applied to the pairs of test resistors $R_T + R_{TT}$ (2 MΩ+100 kΩ) instead of the thermistor. A test signal (DC or dynamic) may be injected at the node between the resistors $R_T$ and $R_{TT}$ which enables calibration of the biasing system including the load resistors $R_{L2}$ and $R_{L3}$.

The reason behind the configuration used for $L_{1A}$, $L_{1B}$ can best be explained by reference to Figure 5, which focuses on one-half of the network connected to the node $V_{bias+}$ of Figure 3. By symmetry, resistor $R_{THE}$ is split into 2 halves and connected to ground.

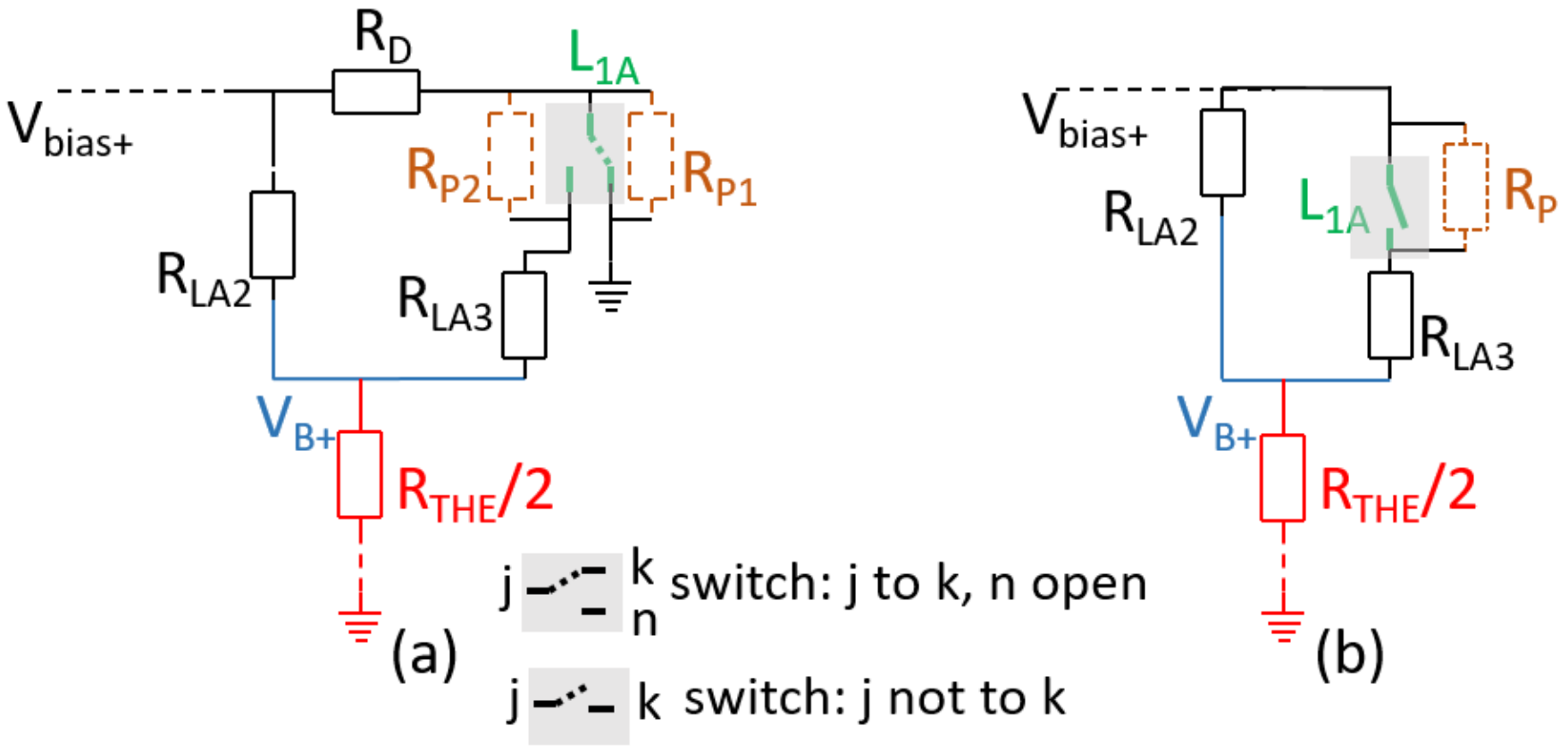

**Figure 5: (a): Detailed view of the node $V_{bias+}$ of the circuit of Figure 3; (b) shows another possible, and more conventional, configuration for $L_{1A}$. For both (a) and (b): by symmetry with respect to Figure 3 thermistor $R_{THE}$ is split into 2 halves and connected to ground.**

Our initial choice was the classical configuration (b) of Figure 5: $R_{LA3}$ in series with $L_{1A}$. We noticed a large thermal drift (a few hundreds of ppm/°C) when $L_{1A}$ was in the open state. The source was drift in $R_P$, the parasitic resistance of the insulating material of the relay. The drift of $R_P$ was reflected in the biasing current because the voltage was applied to $R_{LA2}||(R_{LA3}+R_P)$:

$$\Delta I_B \approx -\frac{V_{bias+}}{R_P}\frac{\Delta R_P}{R_P} \qquad \text{(by considering that } R_P \gg R_{LA3}, R_{THE}) \tag{3}$$

In Figure 5 (a) the parasitic resistance $R_{P2}$ is still present in parallel to the switch in open position, but this time $R_{LA3} + R_{P2}$ is connected between the thermistor and ground, the voltage across $R_{P2}$, is much smaller than $V_{bias+}$, and therefore the drift is greatly attenuated and made negligible:

$$\Delta I_B \approx -\frac{V_{bias+}}{R_{P2}}\frac{1/2\,R_{THE}}{R_{LA2}}\frac{\Delta R_{P2}}{R_{P2}} \qquad \text{(by considering that } R_{P2}, R_{LA2} \gg R_{LA3}, R_{THE}) \tag{4}$$

The microcontroller on the mainboard manages the biasing system. Voltages $V_{bias}$ are read with a 24 bit ΣΔ ADC (AD7732). Based on their values the microcontroller sets the trimmers in order to obtain the final voltage with good symmetry and to compensate for the spread in the load resistor values, if necessary.

### 3.4 The differential preamplifier

The first stage of amplification of the detector signals is at room temperature. Differential inputs and outputs are required to suppress cross-talk between adjacent channels. The wires that connect the thermistors to the first amplifiers are twisted pair for most of their length, and members of a pair are in any case closely spaced to minimize the coupling of ambient fields to differential mode signal. The preamplifier efficiently rejects induced common mode.

The configuration of the preamplifier is similar to an instrumentation amplifier with only one pair of JFET transistors at its input, to minimize noise. The JFETs are custom pairs selected from the NJ132 process from INTERFET. The transistors were selected for a small gate-to-source pinch-off voltage so as to minimize the dropout voltage and thus the gate current [35]. The typical gate current was about 30 fA at room temperature and about 100 fA at 50°C, which is close to the working temperature. The parallel noise developed by the two input JFETs is remarkably low, less than 0.15 fA/√Hz (compared with the 0.5 fA/√Hz attributable to the load resistors). The series white noise is about 3 nV/√Hz rising only to 7 nV/√Hz at 1 Hz, which is smaller than the noise level of the thermistor (thermal noise of 100 MΩ at 10 mK is 7 nV/√Hz). Required fundamental characteristics of the preamplifier are small thermal drift and high gain stability. Although the two JFETs in the pair are not on the same die, the final compensated drift is on average around 0.2 μV/°C.

Figure 6 shows the photograph of the preamplifier. On the right of the figure close to the corners of the board are the two small input connectors.

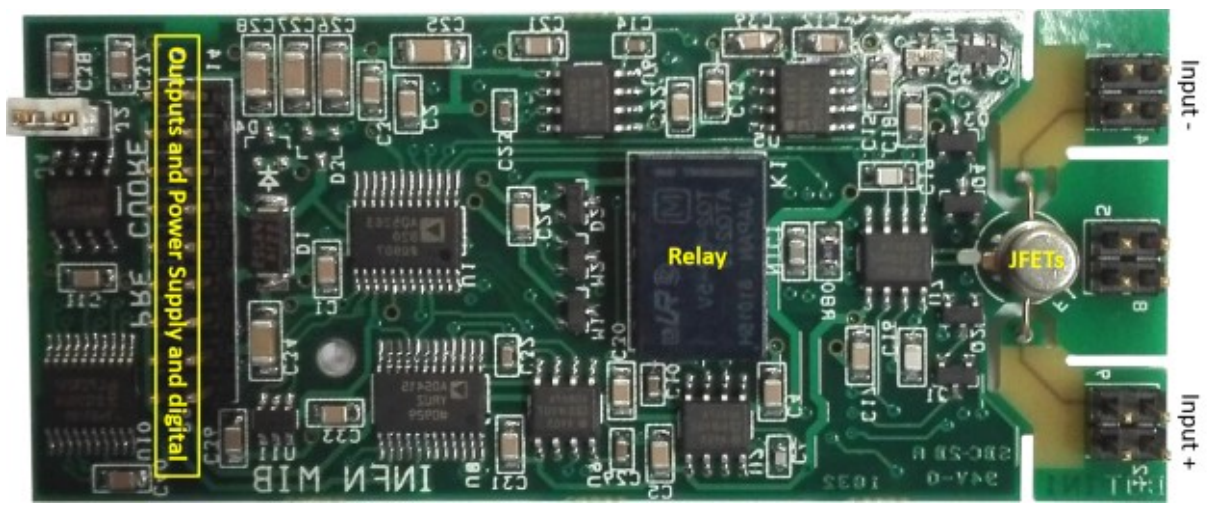

**Figure 6: Photograph of the TOP of the preamplifier, the side that faces the mainboard in Figure 2.**

The package containing the JFETs is close to these two connectors. Its shaped gate leads diverge to the extent possible from the leads of the sources and drains. The gate regions are well isolated from the rest of the circuit to maximize the parasitic impedance. Near the left end of the board is a 16-pin connector, which accommodates power, reference voltages, and digital communication as well as outputs. The present preamplifier evolved from our previous design [36] [37] [38]. Figure 7 and Figure 11 show the complete schematic.

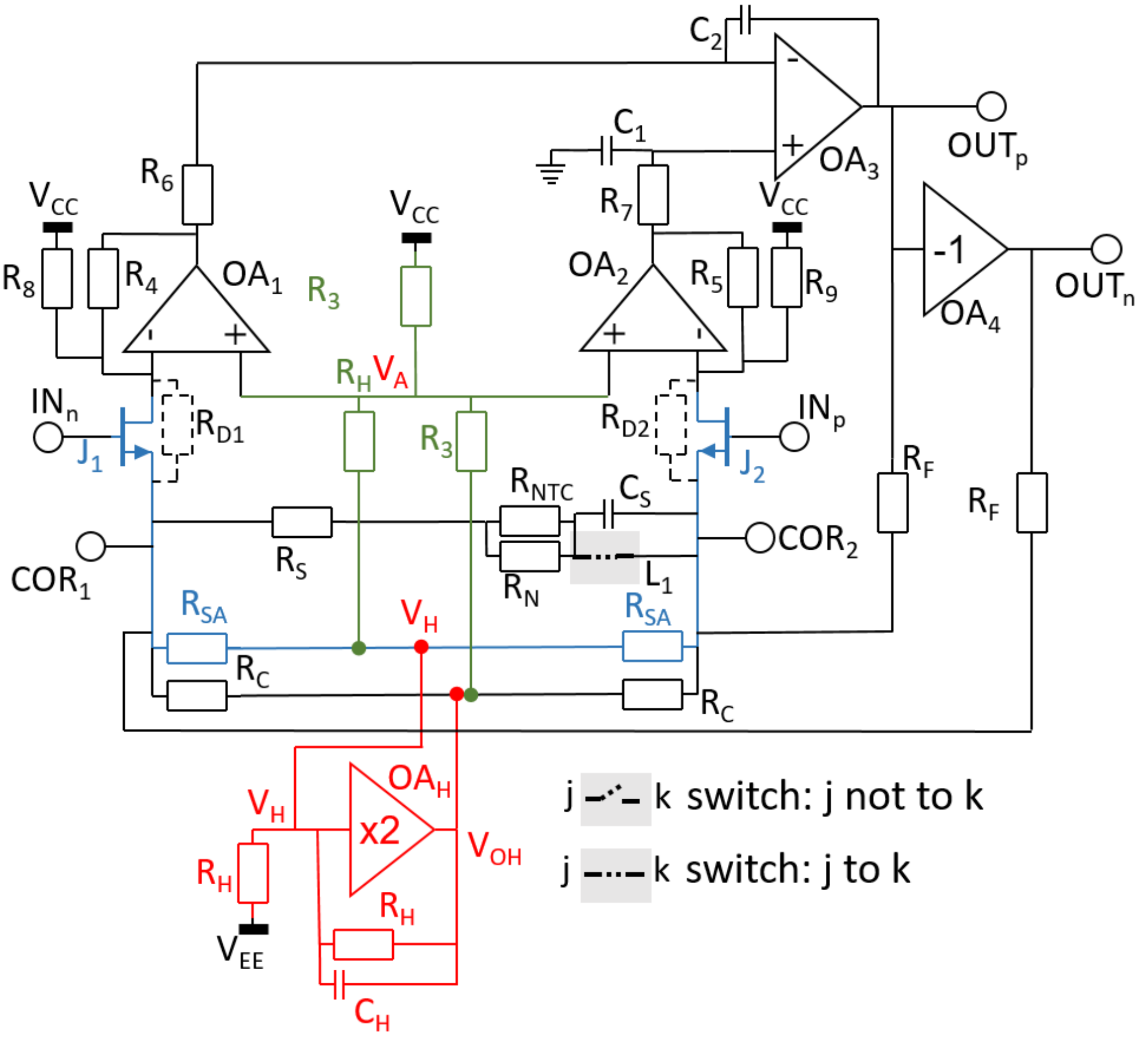

**Figure 7: Schematic diagram of the preamplifier. Nodes $COR_1$ and $COR_2$ are where currents are injected for compensation according to the schematic of Figure 11.**

The aim of the design was to operate the input JFETs, $J_1$ and $J_2$, at constant channel current and channel dropout voltage, to maintain a very stable response under different input signals. This condition helps also in keeping the thermal drift very small after compensation, as it will be shown below. Constant dropout and constant current in JFETs were obtained by maintaining symmetry in the circuit and by using a slightly modified version of the Howland current source, which, notwithstanding its undesirable features [39], proved adequate in our case where the frequency range of the signals is narrow. The Howland current source is the amplifier $OA_H$ and nearby components shown in red in Figure 7. Normally its output is node $V_H$, but we used also the output, $V_{OH}$, of $OA_H$ in the circuit. Let's suppose an equilibrium condition for which $V_{INn}$, $V_{INp}$, are equal. Suppose also that switch $L_1$ is closed, which is its normal operating position. Writing the balance of the currents at node $V_H$, we obtain ($C_H$ is a small compensating capacitance):

$$\frac{V_A - V_H}{R_H} + 2\frac{V_{COR1} - V_H}{R_{SA}} + \frac{V_H}{R_H} - \frac{V_H - V_{EE}}{R_H} = 0 \qquad (V_{COR1} = V_{COR2}) \qquad (5)$$

The original idea of the Howland configuration is the presence of the term $V_H/R_H$, which appears in red in the above equation. This term, which comes from resistor $R_H$ connected between nodes $V_H$ and $V_{OH}$ (considering $V_{OH}=2V_H$), simplifies with the corresponding term on its right in (5), forcing the sum of the first 2 terms on the left to be constant. Next we write current balance at node $V_A$ and again use $V_{OH}=2V_H$:

$$\frac{V_{CC}-V_A}{R_3}+\frac{2V_H-V_A}{R_3}-\frac{V_A-V_H}{R_H}=0 \;\rightarrow\; V_A-V_H=R_H I_{HA}$$
$$\text{where } I_{HA}=\frac{V_{CC}}{2R_H+R_3} \qquad (6)$$

The important result from (6) is that $V_A$-$V_H$ is constant. Substituting $R_H I_{HA}$ for $V_A$-$V_H$ in (5) we obtain

$$\frac{V_{COR1}-V_H}{R_{SA}}=\frac{I_H-I_{HA}}{2} \qquad \text{where } I_H=\frac{-V_{EE}}{R_H} \qquad (7)$$

Thus the current through $R_{SA}$ is also constant. The local feedback of $OA_1$ and $OA_2$ ensures that the drains of $J_1$ and $J_2$ are both at the potential $V_A$. From (7) and (6) we obtain the drain-source voltage $V_A$-$V_{COR1(2)}$:

$$V_A-V_{COR1}=\frac{2R_H+R_{SA}}{2}I_{HA}-\frac{R_{SA}}{2}I_H \qquad (8)$$

which is likewise constant. The high gain, small input offset voltage, and small input offset current of amplifier $OA_3$ ensures near equality of the potentials at the lower terminals of $R_6$ and $R_7$. In turn the currents in $R_4$ and $R_5$ (and in $R_8$ and $R_9$) must be equal and thus also the currents $I_{D1}$ and $I_{D2}$ in the JFETs. $OA_1$ and $OA_2$ thus behave as a cascode current mirror. ($C_1$ and $C_2$ are 4.7 nF compensating capacitances which limit the bandwidth to about 5 kHz).

Let's consider connected two equal resistors $R_{CO}$, $R_{CO1}$ and $R_{CO2}$, whose presence will be justified below in connection to Figure 11, between nodes $COR_1$ and $COR_2$ and ground, then we put $R_C=R_{CO}||R_F$. Writing the current balance for the source terminals of the two JFETs, we obtain:

$$I_{D1(2)}=\frac{I_H-I_{HA}}{2}\left(1+\frac{2R_{SA}}{R_F}\right), \quad \text{and} \quad V_{OUT_P}=V_{OUT_N}=0 \qquad (9)$$

So far, we have shown that, when the inputs are at the same voltage, the currents through and the voltages across the channels of the input JFETs are equal and constant, and the outputs are at zero potential. A differential input signal would determine a change of the output voltage given by (assuming $R_{NTC}>>R_N$):

$$V_{OUT_P}-V_{OUT_N}=2\frac{R_{SE}+R_F}{R_{SE}}\left(V_{COR_2}-V_{COR_1}\right), \quad R_{SE}=(R_S+R_N)||2R_{SA}||2R_C \qquad (10)$$

In this case the drain-source voltages of $J_1$ and $J_2$ become different by $\left(V_{COR_1}-V_{COR_2}\right)$, while their currents remain equal.

The currents through the drain resistors $R_{D1}$ and $R_{D2}$ change when the drain-to-source voltage changes. Because the current through the JFET is constant, it follows that:

$$V_{COR_{1(2)}}=\frac{g_{m1(2)}R_{D1(2)}}{1+g_{m1(2)}R_{D1(2)}}V_{IN_{p(n)}} \qquad (11)$$

where $g_{m1(2)}$ is the transconductance of the JFET. Therefore

$$V_{OUT_P}-V_{OUT_N}=2\frac{R_{SE}+R_F}{R_{SE}}\frac{g_{m1(2)}R_{D1(2)}}{1+g_{m1(2)}R_{D1(2)}}\left(V_{IN_p}-V_{IN_n}\right),$$
$$R_{SE}=(R_S+R_N)||2R_{SA}||2R_C \qquad (12)$$

With $g_m$ of roughly 5 mA/V and $R_{DS}$ roughly 12 kΩ the factor dependent on $g_m R_D$ is close to one. We added the NTC resistor $R_{NTC}$ in parallel with a 5 Ω metal film resistor to cancel the drift in voltage gain that would result from thermal drift in the JFET parameters. With $R_F$ equal to 20 kΩ and $R_{SE}$ 190 Ω the differential gain is about 210 V/V. The channel current of each JFET is set at 0.5 mA, and $V_{DS}$ is 0.5 V. Figure 8 shows a histogram of the drift of the gain in the temperature range from 25 °C to 35 °C. The mean is 3.1 ppm/°C and the sigma is 2.8 ppm/°C.

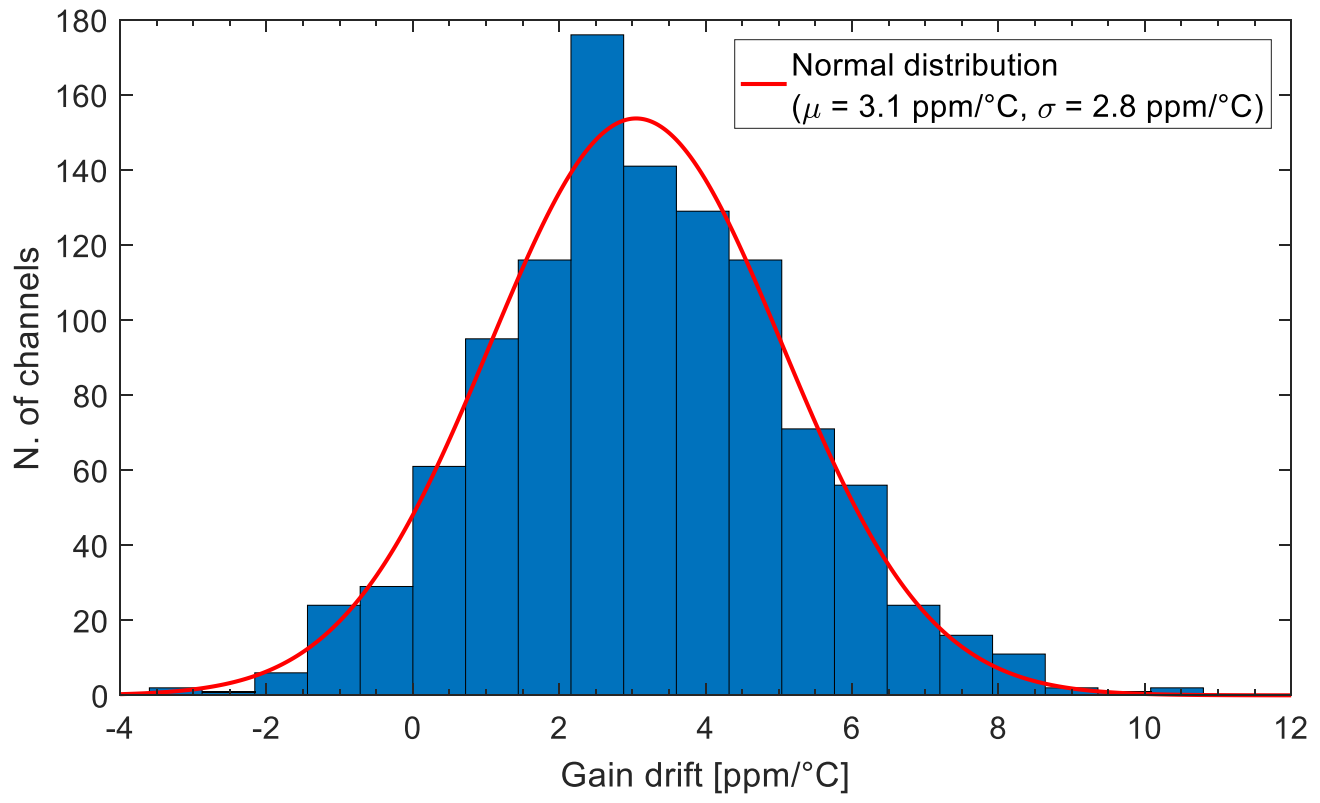

**Figure 8: Distribution of gain drift of preamplifiers over the temperature range 25 °C to 35 °C. The average value is (μ) 3.1 ppm/°C and the standard deviation (σ) is 2.8 ppm/°C.**

For every channel on every fully configured mainboard, i.e. preamplifier followed by programmable gain amplifier, we measured white noise and noise in vicinity of 1 Hz. Figure 9 and Figure 10 show the results. For white noise the average across channels was 3.2 nV/√Hz referred to the input, and at 1 Hz the average was 7.8 nV/√Hz. Figure 10 shows the noise spectrum of a typical channel. The noise rolls off above 5 kHz because of the passband limit designed into the preamplifier.

We set switch $L_1$ of Figure 7 to the open position for load curve characterization. In this configuration the DC gain is reduced to 27 V/V, and the preamplifier dynamic range is extended to accommodate large thermistor bias. At the lower gain stability can be a concern, but capacitor $C_S$ (4.7 nF) ensures that the gain at high frequency remains unaffected.

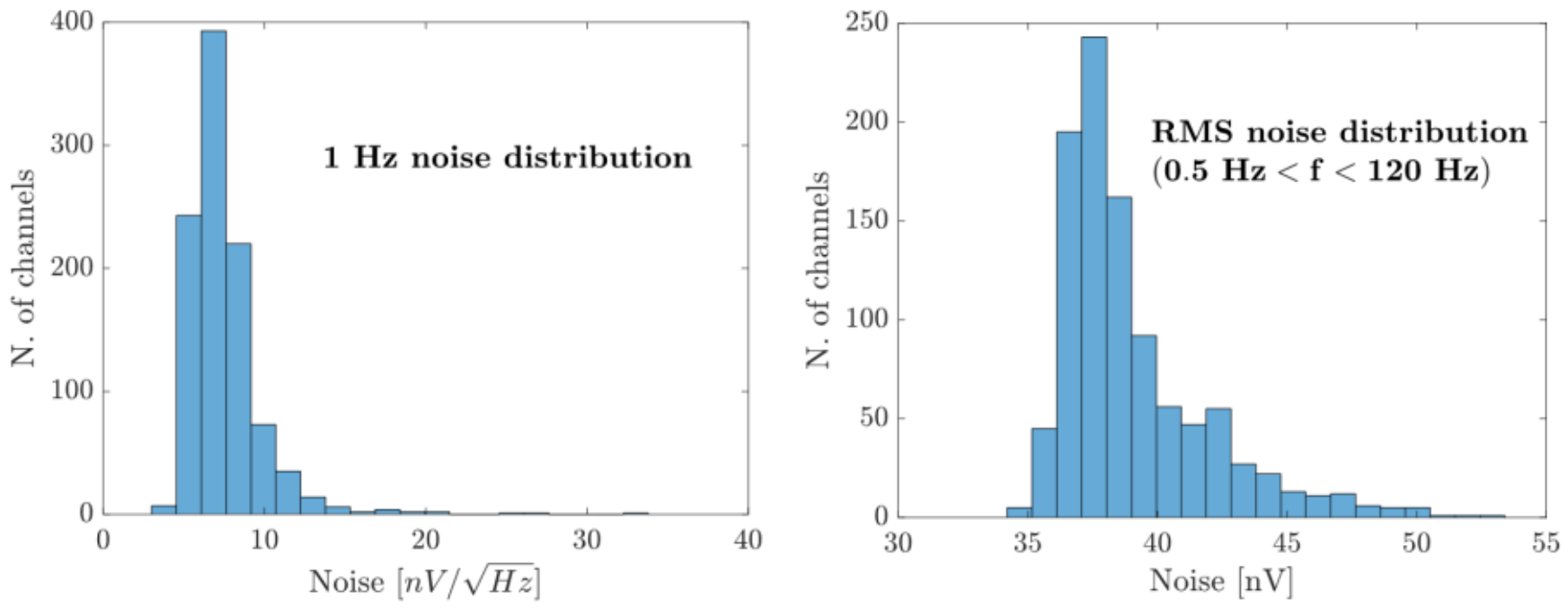

**Figure 9: Distribution of series noise of the 1000 preamplifiers. (a) Noise at 1 Hz. The centroid of histogram is 7.5 nV√Hz. (b) RMS noise in the range 0.5 Hz - 120 Hz. The centroid of the histogram is 39 $nV_{RMS}$.**

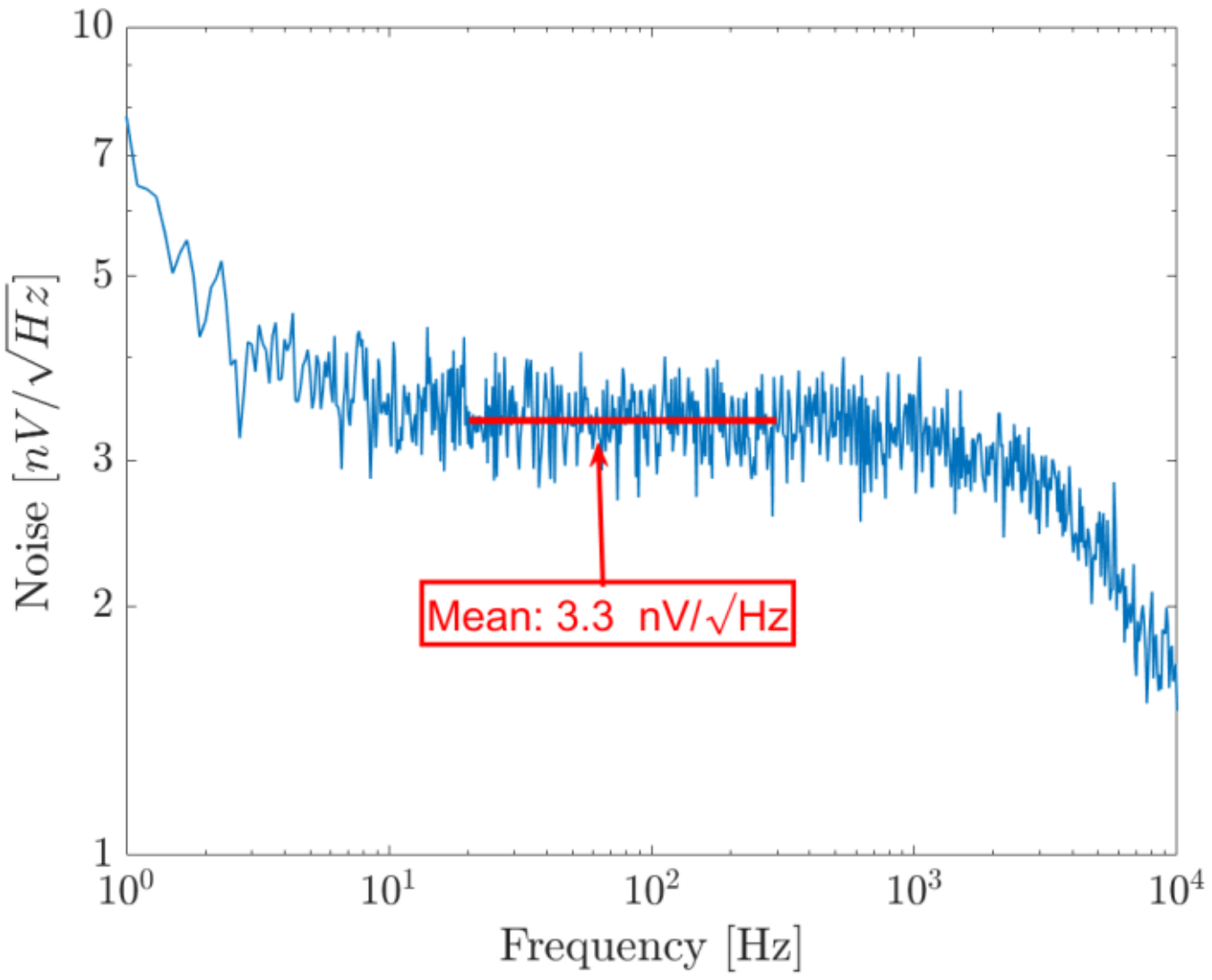

**Figure 10: Typical input series noise of the preamplifier/PGA chain. Note the frequency bandwidth limited to about 5 kHz.**

The preamplifier description above presumes perfect matching between components, which is far from reality particularly for the JFET pair. In practice thermal drift and input offset are present, and we are able to compensate for both with the additional circuit of Figure 11.

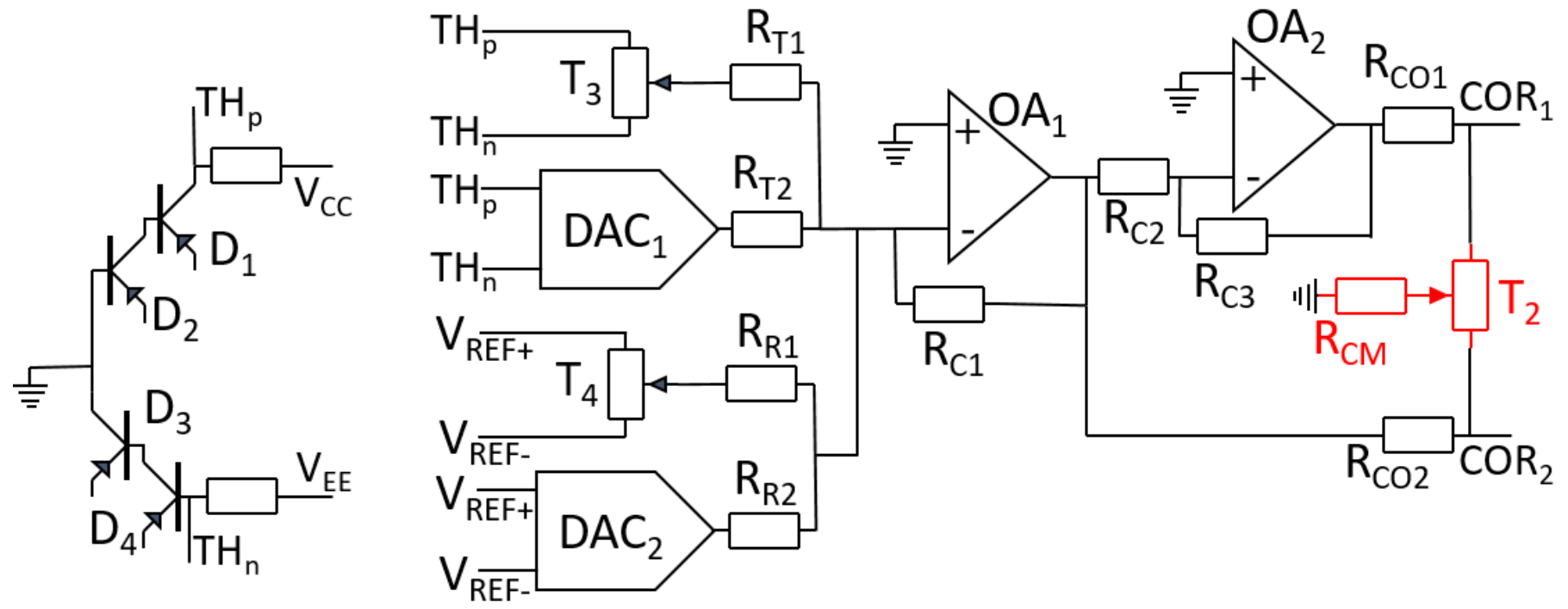

**Figure 11: Drift, thermal and common mode correction circuitry for the preamplifier of Figure 7. $DAC_1$ and $DAC_2$ negative references are connected- to an internal available resistor, see datasheet.**

Offset voltage is adjusted with the 8-bit digital trimmer $T_4$ (1/4 of AD5263) and the 12-bit current output $DAC_2$ (1/2 of an AD5415). The voltage at the wiper terminal of $T_4$ is multiplied by the ratio $R_{C1}/R_{R1}$ which is close to unity. The output of $OA_1$ is mirrored with $OA_2$ and the differential pair is applied via resistors $R_{CO1}$ and $R_{CO2}$ and nodes $COR_1$ and $COR_2$ to the sources of the JFETs (see Figure 7). The output of $DAC_2$ follows the same path, but because the resistance $R_{R2}$ is about 15 times larger than the resistance $R_{R1}$, the weight of $DAC_2$ is 15 times smaller than that of $T_4$. The reference voltage $V_{REF+}$ for $T_4$ and $DAC_2$ comes from our stable, low-noise ±5 V voltage generator [19]. $V_{REF-}$, on the other hand, is tied to a resistor internal to the AD5415 (see the AD5415 datasheet). The maximum input compensation is about ±60 mV, more than enough to both compensate for the ±20 mV maximum offset between JFETs and the bias voltage applied to the detector. Equal resistors $R_{CO1}$ and $R_{CO2}$ connect the circuit of Figure 11 to nodes $COR_1$ and $COR_2$ of Figure 7, justifying the reason we set $R_C$'s of Figure 7 equal to $R_F||R_{CO}$ for maintaining symmetry.

The network for thermal drift compensation works through the same path as the offset correction. In this case the adjustable currents come from trimmer $T_3$ and $DAC_1$. The potential across the stacks $D_1$, $D_2$, and $D_3$, $D_4$, of forward biased diodes serve as the $TH_p$ and $TH_n$ reference inputs. The potential of the stacks has a temperature dependence of ±4 mV/°C. This range enables compensation at the preamplifier input up to about ±40 µV/°C with an accuracy of about 100 nV/°C.

Drift compensation requires calibration, which was performed crate by crate in an environmental chamber that established an ambient temperature for the mainboards in the range 20 °C ÷ 50 °C. In a first run a coarse coefficient for $T_3$ was determined. Using this coefficient we are then able to set the drift close to 0.5 µV/°C, irrespective of temperature and thermistor bias current. In a second run the fine value for $DAC_1$ was determined as a linear function of the $T_4$ and $DAC_2$ settings. Residual temperature dependence of the amplifier offset, however, implied quadratic dependence and the need for greater sophistication. In consequence we split the second run into three temperature regions, 20 °C ÷ 30 °C, 30 °C ÷ 40 °C and 40 °C ÷ 50 °C, and determined a fine coefficient for each region. Finally, a last run within the environmental chamber is done for verification. This automated procedure takes about 3 days per crate, as the temperature is slewed very slowly. Figure 12 shows the distribution of the drift with no compensation, with only coarse compensation, and also with fine-tuned compensation.

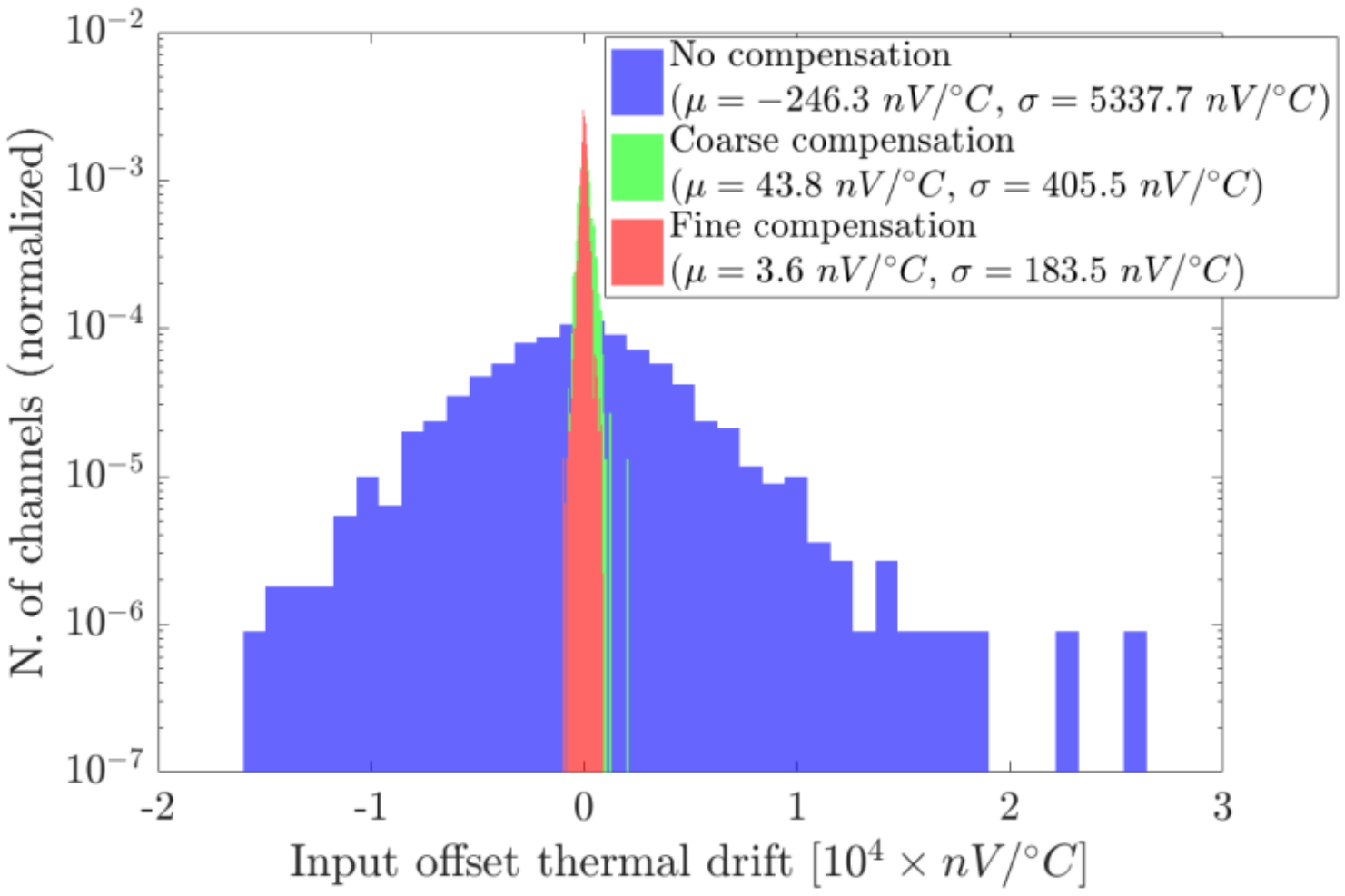

**Figure 12: Production distribution of the preamplifier input voltage drift before calibration and after coarse and fine calibration. Vertical scale is logarithmic. Temperature range is from 20 °C to 50 °C. In all cases the average value of the histograms is close to zero (µ), while the RMS distribution in the uncompensated condition is (σ) about 5 µV/°C, after the coarse compensation 0.4 µV/°C and, after the final fine compensation, 0.18 µV/°C.**

The RMS distribution after fine calibration is less than 190 nV/°C over the full temperature range spanning 20 °C to 50 °C. The measured coefficients are stored in an EPROM located on the preamplifier board. The output offset is adjusted following one of two protocols: fast adjustment when only the coarse coefficient is used, or slow adjustment, which is implemented in two steps. In the first step only the coarse coefficient is set and the offset is adjusted and the temperature measured (by an LM73 thermometer located on every preamplifier). Based on the resulting values for $T_4$, $DAC_2$, and the temperature, the proper coefficients are loaded from the EPROM and used to calculate the value for $DAC_1$.

The common mode rejection ratio, CMRR, is optimized with trimmer $T_2$ and resistor $R_{CM}$ (1 MΩ) shown in Figure 11. We apply an asymmetric bias to the input and adjust $T_2$ to minimize the output voltage. Figure 13 shows that after tuning the CMRR is never smaller than 120 dB at DC.

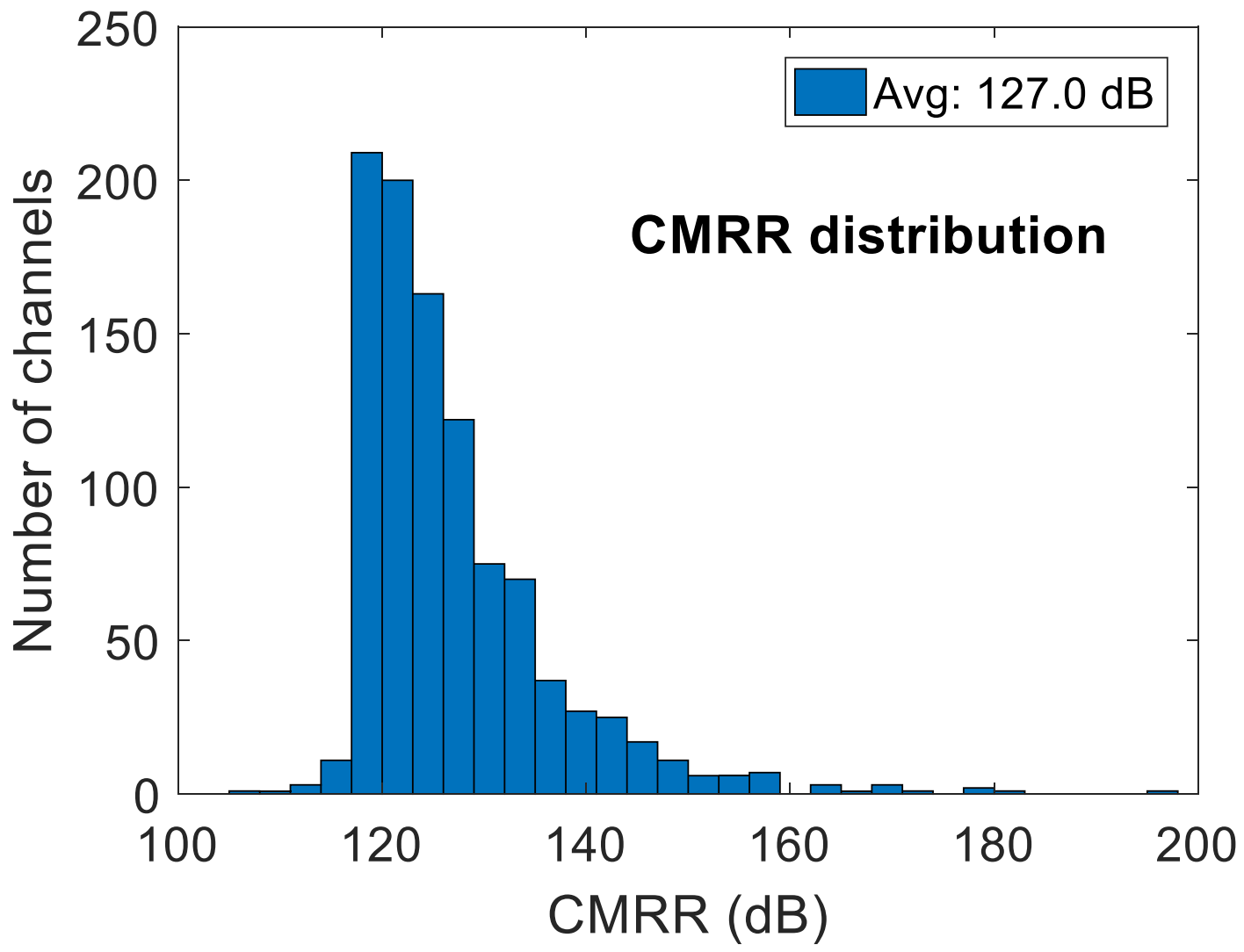

**Figure 13: Common Mode Rejection Ratio distribution of the full production after optimization of trimmer $T_2$ of Figure 11.**

Good rejection is important not only for the suppression of common mode disturbances, but also to suppress parallel noise from the load resistors. If we call $A_d$ the differential gain and $A_c$ the common mode gain of the amplifier of Figure 14, then at the amplifier output the noise contribution from the load resistors only is:

$$\overline{V_O^2} = A_d^2 \left[ \frac{\overline{e_L^2}}{2R_L^2} R_{THE}^2 + \left( \frac{A_c}{A_d} \right)^2 \frac{\overline{e_L^2}}{2} \right] \tag{13}$$

If we require to the second term, coming from the common mode amplification, to be lower than 1 nV/√Hz, the ratio $A_c/A_d$ should be higher than 84 dB for our 30 GΩ load resistors.

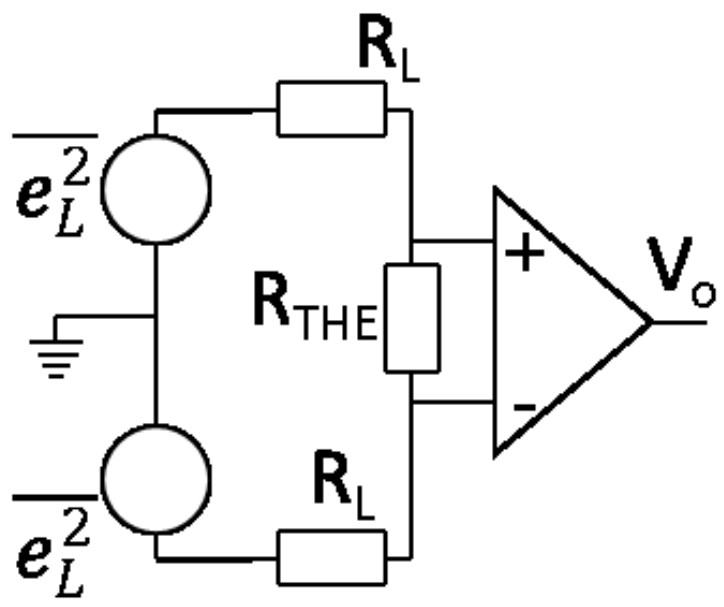

**Figure 14: Model for the parallel noise at the preamplifier inputs.**

### 3.5 Second stage Programmable Gain Amplifier

The second stage amplifier is a programmable gain amplifier, PGA, settable from 1 V/V to 50 V/V. This range of adjustability is a necessity related to the great spread of energy gain of the detectors. Its schematic circuit is shown in Figure 15. At the input is a differential preamplifier whose coarse gain can be set to 10 V/V, SWs open, or 1 V/V, SWs closed.

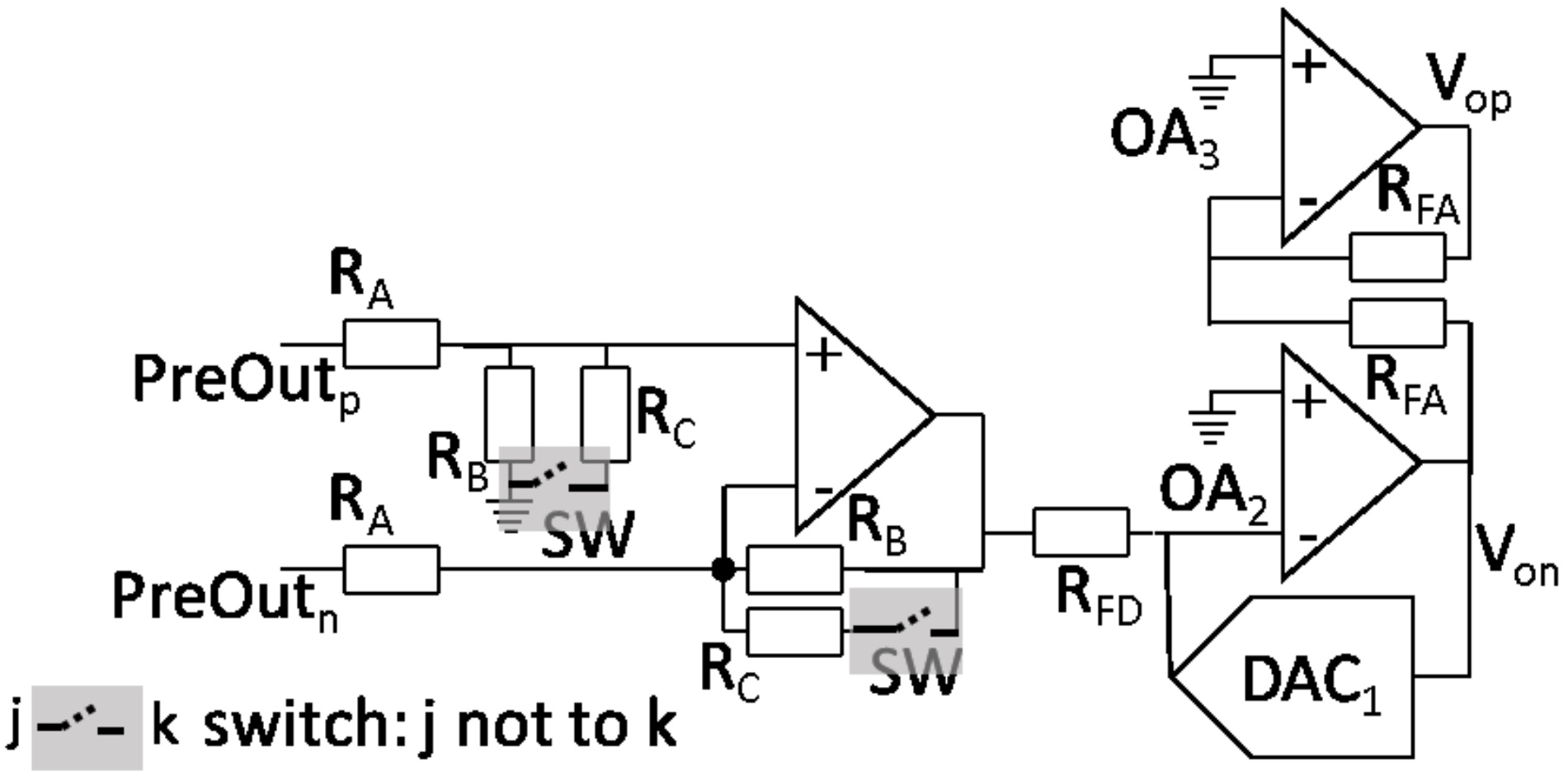

**Figure 15: Second stage PGA.**

We set the fine multiplier gain factor, from 1 V/V to 5 V/V, by means of the DAC in the feedback path of $OA_2$ (see AD5415 datasheet). The output is made differential by $OA_3$ in unitary inverting gain. Resistors $R_{FA}$ and $R_{FD}$ are all internal to the DAC, with a relative stability of a few ppm/°C, while resistors $R_A$, $R_B$, and $R_C$ are metal film. The gain stability of this stage is better than 5 ppm/°C.

### 3.6 Slow control

The FE board has a number of features that need to be set, and a microcontroller, a 32-bit ARM LPC2468, is dedicated to this purpose. It operates with the support of a number of digital chips and two 24-bit ΣΔ-ADC (AD7732). As shown in Figure 2, most of these digital circuits are located close to the output connectors, i.e. as far as possible from the inputs, to minimize noise injection. The microcontroller is managed by the DAQ system. It controls the mainboard according to instructions issued from the DAQ. The communication protocol adopted is the addressable serial CAN-bus.

Several functions are available, and here we will discuss only the most significant ones. Through a set of bi-stable relays power may be applied to or withheld from the six channels of a board individually. Some relay drivers (MAX4820) are operated by the microcontroller through Serial Peripheral Interface, SPI. The SPI is also used to set the DACs (PGA gain set). The digital trimmers (detector bias) and the preamplifier are in communication with the microcontroller by means of I2C, Inter-Integrated Circuit, serial bus.

The most important and critical functions are the setting of the output offset voltage and the thermistor bias current. The former can be adjusted following various procedures. In the case that the baseline exhibits large fluctuations the standard Successive Approximation Register technique, SAR, is preferred because it finds the final guess by trial and test. A faster and more effective procedure implements SAR until the output voltage saturates at one rail. After restoring the output to the dynamic operating range, the final voltage is calculated from the known slope (voltage change per bit of preamplifier trimmer – preamplifier DAC). This guess procedure can be repeated if the result of the first try is outside a specified tolerance. During this procedure the PGA gain is set to 1 V/V. During the offset procedure, adjustment of the temperature drift corrector as well is an option.

Whenever the offset is adjusted following a planned gain change, an algorithm a) measures the output offset prior the gain change, b) sets the PGA gain to 1, calculates and updates settings for the preamplifier trimmer and DAC that will restore the output voltage to the status quo ante following the gain adjustment, and c) sets the new gain. This algorithm is designed to prevent as much as possible sudden saturation of the outputs because saturation may predispose the channel to experience cross-talk.

The approach to setting the thermistor bias current is similar to the method for the offset correction, but in this case preservation of symmetry is an additional consideration. The biasing is symmetric when $V_{B-} = -V_{B+}$ (see Figure 3). In addition, during the setting of the bias current the overall gain is set at the minimum value and the output offset is set to zero to minimize disturbances coming from sudden changes. After the bias is set, the gain is restored to its original value. Either the offset correctors may be adjusted to restore the original offset, or the preamplifier trimmer and DAC may be restored to their previous values. The second option reveals the impact of the biasing change on the preamp output, and is exploited during the detector characterization phase.

To minimize the number of digital lines the on-board digital communication is based on serial buses SPI and I2C. Chip selection, where needed, is implemented in a master-slave arrangement. The slaves are multiplexers and decoders (examples are ADG408, SN74CB3T3245, PCA9554), which are geographically close to the chips they control. The microcontroller manipulates the slaves via a serial bus. All the digital lines are embedded between two ground layers and respect a generous stay-clear with respect to the inputs of the board. When the microcontroller is not active, it enters a power-down state, and all clock signals become quiescent. In this condition remaining power requirements in the digital section are met from the analog power supply. When a remote command awakens the microcontroller, the power source for all the digital circuitry is shifted to a dedicated external line.

## 4. Anti-aliasing filter

The DAQ acquires continuously: triggering and filtering is performed in software. Before digitization we have an anti-aliasing active filter (Bessel-Thomson type) that features a roll-off of 120 dB/decade and a -3 dB bandwidth selectable from 15 Hz up to 120 Hz in four steps [40]. The gain below the cutoff frequency is unity. The thermal noise of the resistors needed in the bandwidth selection limits the input series noise and is about 60 nV/√Hz for the 15 Hz bandwidth where the larger noise is expected. The input RMS thermal drift is about 1.5 μV/°C as measured in the 40 °C to 60 °C temperature range as shown in Figure 16. Each filter board has 12 channels and a microcontroller, an ARM LPC2129, manages their operation following remote instructions from the DAQ.

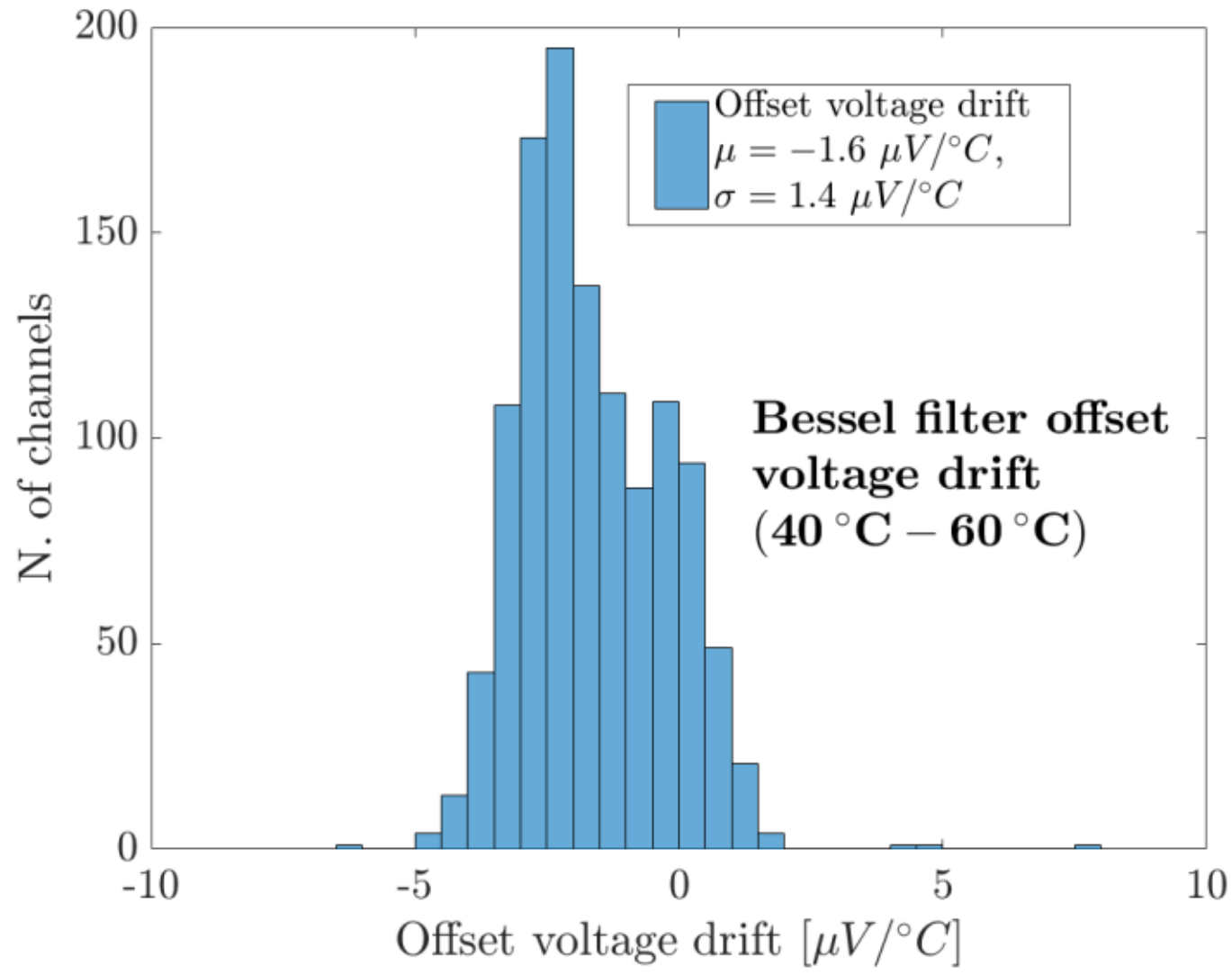

**Figure 16: Distribution of the input thermal drift of the anti-aliasing filter. The average of the drift over all the channel is close to -1.6 μV/°C, while the RMS distribution is 1.4 μV/°C.**

## 5. Power supply

The power dissipation of all of the front-end electronics inside the Faraday cage is about 700 W. The power supplies have been developed following the Point Of Load approach, POL. In this paradigm the final stage of power generation for a load is placed geographically close to the load to facilitate optimal regulation and stability. Figure 17 shows the scheme we used.

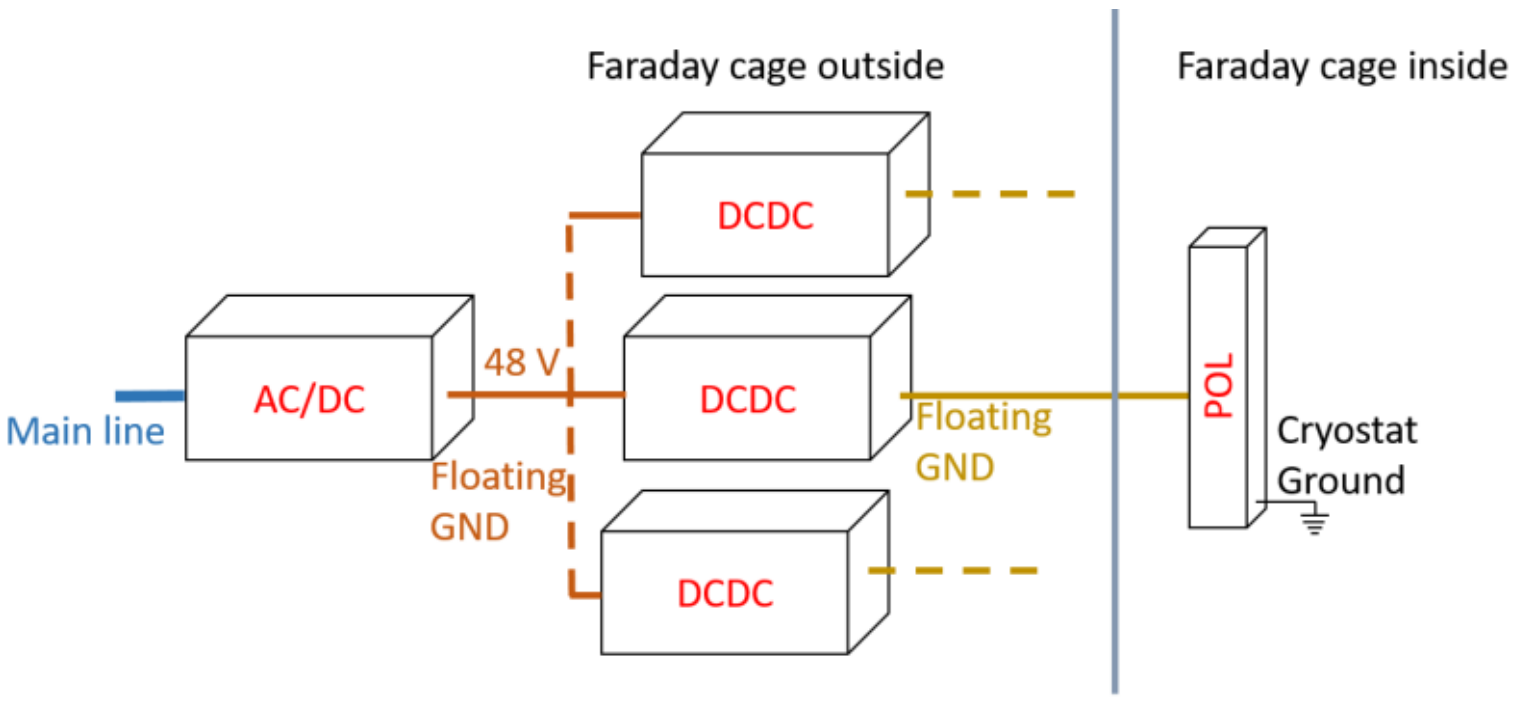

**Figure 17: Diagram of the power supply system.**

A commercial AC/DC power supply provides power at 48 V to several DC/DC units [13] each with two analog outputs adjustable between ±5 V and ±15 V, and one digital output adjustable from 5 V to 15 V. The DC/DC units power a constellation of POL power supplies embedded in the Front-end crates. The POL PS is a linear power supply. It generates ±5 V with excellent regulation, noise less than 50 nV/√Hz down to 1 Hz, and stability at the level of 1 ppm/°C [19]. This performance is good enough that we are able to use the output not only for power but also as a voltage reference in the offset adjustment and thermistor biasing circuitry.

A pair of POL PS supplies a crate and a DCDC feeds the pair. To supply the overall Electronics system we used 4 ACDCs, 20 DCDCs (with one spare) and 28 POL PSs.

The use of floating grounds on both the AC/DC and DC/DC regulator outputs enables the creation of a single ground node at the cryostat, which confers a degree of immunity to ground loops. Common grounding of the detector as a whole, which cannot be left floating for safety reasons, is at the DAQ input stage (Figure 1).

## 6. Communication system

The communication links between the DAQ, the mainboards, and the anti-aliasing filters are critical because of the risk that they inject noise and create ground loops. To mitigate these risks we have modified the standard connection of the CAN-bus serial link by introducing fibre optics on the main paths. Figure 18 shows the classical configuration in which most devices connect to the CAN-bus via a transceiver that mediates between unidirectional transmitter and receiver lines, TD/RD, on the device side and a single bi-directional, differential line on the bus side.

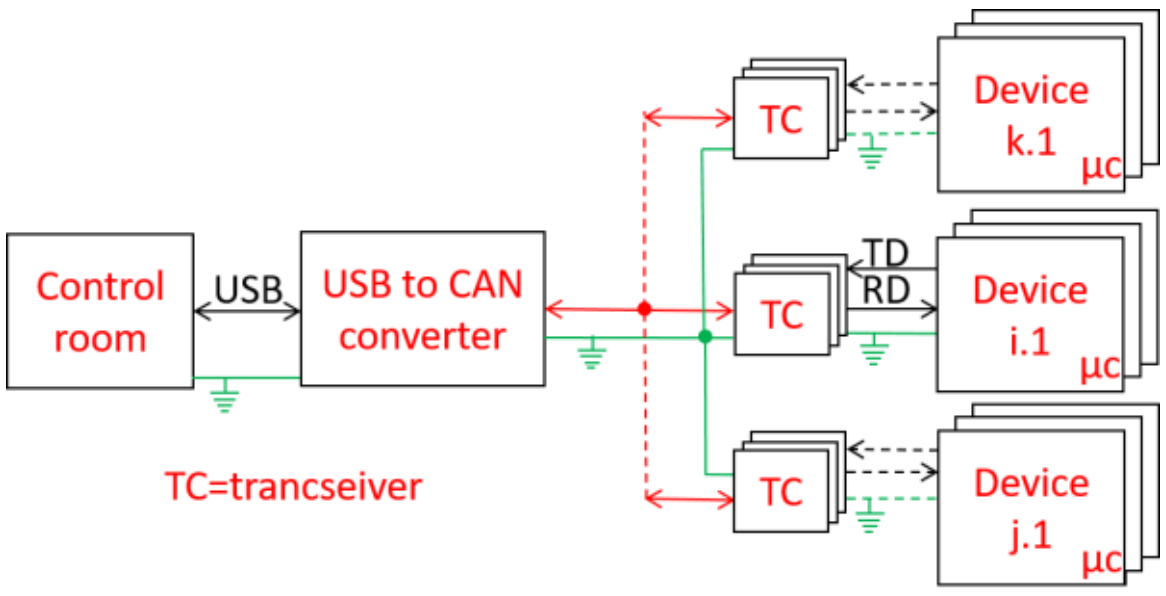

**Figure 18: Standard CAN-bus connections.**

The computer connects via a "USB to CAN-bus converter." In this configuration devices connected to the network have a ground reference in common. A ground loop is likely to be present, and, moreover, the objectionable ground of the control room computer reaches the mainboards, detectors, etc.

We avoided the ground connections by modifying the links as in Figure 19.

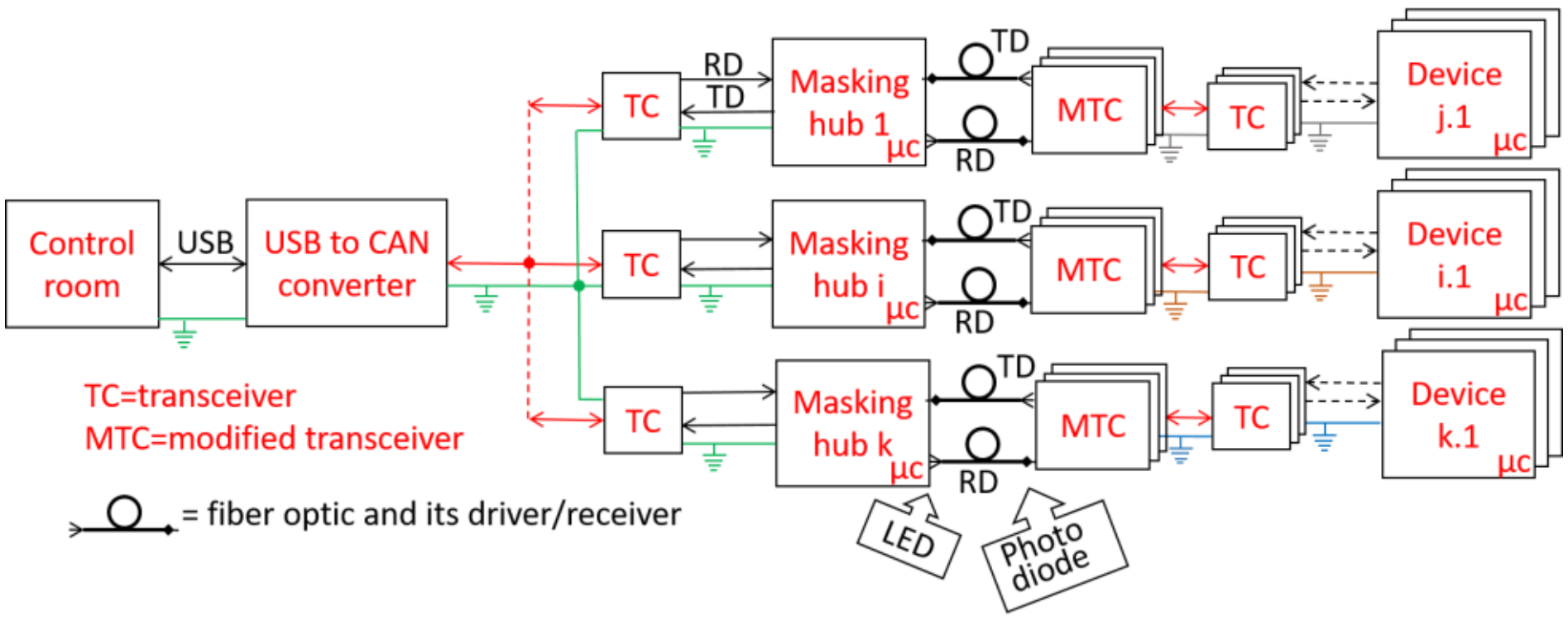

**Figure 19: CAN-bus connecting system based on fibre optic transmission: the ground of the input is isolated from the outputs, thanks to the presence of the fibre optics, and every output has the ground isolated each other, avoiding ground loop from this side.**

In addition we organized the system into homogeneous groups. A message from the "USB to CAN converter" arrives at a series of ordinary transceivers, TCs. The TCs convey the message to the TD and RD lines, which communicate with "masking hubs." A message continues beyond a masking hub only if its address field includes the sub-address belonging to the masking hub itself. Thus only the target sub-group receives the message and wakes up. The other sub-groups remain free of unnecessary digital activity and concomitant noise in the front end. The masking hub targeted by the message sends it across a fibre optic link, which facilitates the decoupling of grounds. At the other end of the fibre optic link an "MTC," a modified fibre optic transceiver, converts the message to the bidirectional differential format understood by TCs. A TC dedicated to the sub-group then receives the message from the MTC and forwards it on TD/RD lines to the target devices. What appears to be unnecessary complexity in this scheme arises in part because the conversion between optical and electrical representations of the message is most conveniently accomplished in a segment where signal propagation is unidirectional, i.e. at TD/RD. The blocks in Figure 19 from the control room to the input side of the fibres are all outside the Faraday cage. The rest is inside. This arrangement helps further to reduce the opportunity for noise generation close to the mainboards, as the grounds of the sub-groups are all separated. The masking hubs incorporate an ARM Cortex-M3 microcontroller, LPC1768, as the key element. All messages are buffered with direct memory access, DMA, which helps to prevent corruption of the data streams.

## 7. Summary

We have given a detailed description of the front-end system for the CUORE and CUPID-0 detectors. It consists of front-end boards hosting thermistor biasing circuitry and voltage amplifiers with low noise and low drift. The outputs of the front-end boards are processed by anti-aliasing filters. All boards are controlled remotely via optical links to a slow control system. We have also described the purpose-built chain of power conditioning modules that furnish low-noise power to the signal processing components. These systems now serve both CUORE and CUPID-0 in their data-taking phase.

## 8. Acknowledgements

The CUORE and LUCIFER/CUPID-0 Collaborations thank the Director and staff of the Laboratori Nazionali del Gran Sasso and the technical staff of our laboratories.

This work was supported by the Istituto Nazionale di Fisica Nucleare (INFN) and by the U.S. NSF Division of Physics.

This work also received support, from the European Research Council under the European Union's Seventh Framework Programme (FP7/2007-2013)/ERC Grant Agreement no. 247115. This award was made in the context of the LUCIFER project.

## 9. References

[1] E.Fiorini, A.Pullia, G.Bertolini, F.Cappellani, G.Restelli, A search for lepton non-conservation in double beta decay with a germanium detector, Physics Letters, 25B, 602, 1967.

[2] R.K.Bardin, P.J.Gollon, J.D.Ullman, C.S.W'u, Double beta decay in $^{48}$ca and the conservation of leptons, Physics Letters, 26B, 112, 1967.

[3] E.Fiorini, A.Pullia, G.Bertolini, F.Cappellani, G.Restelli, An Underground Experiment on Neutrinoless Double Beta-Decay, Lettere Al Nuovo Cimento, 3, 149, 1970.

[4] E.Fiorni, T.O.Niinikoski, Low-temperature calorimetry for rare decays, Nuclear Instruments and Methods in Physics Research 224, 83-88, 1984.

[5] S.H.Moseley, J.C.Mather, D.McCammon, Thermal detectors as x-ray spectrometers, Journal of Applied Physics, 56, 1257, 1984.

[6] D.McCammon, S.H.Moseley, J.C.Mather, R.F.Mushotzky, Experimental tests of a single-photon calorimeter for x-ray spectroscopy, Journal of Applied Physics 56, 1263, 1984.

[7] D.McCammon, M.Juda, J.Zhang, S.S.Holt, R.L.Kelley, S.H.Moseley, A.E.Szymkowiak, Thermal Detectors for High Resolution Spectroscopy, Japanese Journal of Applied Physics, 26, 1987.

[8] D. McCammon, M. Juda, and J. Zhang, R. L. Kelley, S. H. Moseley, and A. E. Szymkowiak, Thermal detectors for high resolution spectroscopy, IEEE Transactions on Nuclear Science, 33, No. 1, 236, 1986.

[9] C.Arnaboldi, F.T.Avignone III, J.Beeman, M.Barucci, M.Balata, C.Brofferio, C.Bucci, S.Cebrian, R.J.Creswick, S.Capelli, L.Carbone, O.Cremonesi, A.de Ward, E.Fiorini, H.A.Farach, G.Frossati, A.Giuliani, D.Giugni, P.Gorla, E.E.Hallerj, I.G.Irastorza, R.J.McDonald, A.Morales, E.B.Norman, P.Negri, A.Nucciotti, M.Pedretti, C.Pobes, V.Palmieri, M.Pavan, G.Pessina, S.Pirro, E.Previtali, C.Rosenfeld, A.R.Smith, M.Sisti, G.Ventura, M.Vanzini, L.Zanotti, CUORE: a cryogenic underground observatory for rare events, Nuclear Instruments and Methods in Physics Research, A518, 775, 2004.

[10] J.W.Beeman, F.Bellini, P.Benetti, L.Cardani, N.Casali, D.Chiesa, M.Clemenza, I.Dafinei, S.Di Domizio, F.Ferroni, A.Giachero, L.Gironi, A.Giuliani, C.Gotti, M.Maino, S.Nagorny, S.Nisi, C.Nones, F.Orio, L.Pattavina, G.Pessina, G.Piperno, S.Pirro, E.Previtali, C.Rusconi, M.Tenconi, C.Tomei, and M.Vignati, Current Status and Future Perspectives of the LUCIFER Experiment, Advances in High Energy Physics Volume 2013, Article ID 237973, 2013.

[11] C.Ligi, C.Alduino, F.Alessandria, M.Biassoni, C.Bucci,·A.Caminata, L.Canonica, L.Cappelli, N.I.Chott, S.Copello,·A.D'Addabbo, S.Dell'Oro, A.Drobizhev, M.A.Franceschi,·L.Gladstone, P.Gorla, T.Napolitano, A.Nucciotti, D.Orlandi,·J.Ouellet, C.Pagliarone, L.Pattavina, C.Rusconi, D.Santone,·V.Singh, L.Taffarello, F.Terranova, S.Uttaro, The CUORE Cryostat: A 1-Ton Scale Setup for Bolometric Detectors, Journal ofLow Temp Phys, 184, 590, 2016.

[12] C.Bucci, P.Carniti, L.Cassina, C.Gotti, A.Pelosi, G.Pessina, M.Turqueti, S.Zimmermann, The Faraday room of the CUORE Experiment, Journal of Instrumentation , JINST, 12, P12013, 2017.

[13] C.Arnaboldi, A.Baù, P.Carniti, L.Cassina, A.Giachero, C.Gotti, M.Maino, A.Passerini, G.Pessina, Very low noiseAC/DC power supply systems for large detector arrays, Review Of Scientific Instruments, V 86, p 124703, 2015.

[14] S.Di Domizio, Search for double beta decay to excited states with CUORICINO and data acquisition system for CUORE, thesis, http://inspirehep.net/record/1615525.

[15] C.Alduino et al, CUORE-0 detector: design, construction and operation, Journal of Instrumentation, JINST, 11, P07009, 2016.

[16] A.Alessandrello, C.Brofferio, C.Bucci, O.Cremonesi, A.Giuliani, B.Margesin, A.Nucciotti, M.Pavan, G.Pessina, E.Previtali, M.Zen, Methods for response stabilization in bolometers for rare decays, Nuclear Instruments and Methods in Physics Research, Section A, 412, 454, 1998.

[17] C.Arnaboldi, G.Pessina, E.Previtali, A programmable calibrating pulse generator with multi-outputs and very high stability, IEEE Transaction on Nuclear Science, 50, 979, 2003.

[18] P.Carniti, L.Cassina, A.Giachero, C.Gotti, G.Pessina, A High Precision Pulse Generation and Stabilization System for Bolometric Experiments, arXiv:1710.05565 [physics.ins-det] and submitted to JINST on xx yyy 2017.

[19] P.Carniti, L.Cassina, C.Gotti, M.Maino, G.Pessina, A low noise and high precision linear power supply with thermal foldback protection, Review Of Scientific Instruments, V 87, p 054706, 2016.

[20] A.Giachero, C.Gotti, M.Maino, G.Pessina, Modelling high impedance connecting links and cables below 1 Hz, Journal of Instrumentation, JINST 7, P08004, 2012.

[21] A.Giachero, C.Gotti, M.Maino, G.Pessina, Characterization of high impedance connecting links for bolometric detectors, Nuclear Instruments and Methods in Physics Research A, 718, 229, 2013.

[22] E.Andreotti, C.Arnaboldi, M.Barucci, C.Brofferio, C.Cosmelli, L.Calligaris, S.Capelli, M.Clemenza, C.Maiano, M.Pellicciari, G.Pessina, S.Pirro, The low radioactivity link of the CUORE experiment, Journal of Instrumentation, Jinst, 4, P09003, 2009.

[23] C.Brofferio, L.Canonica, A.Giachero, C.Gotti, M.Maino, G.Pessina, Electrical characterization of the low background cu-pen links of the CUORE experiment, Nuclear Instruments and Methods in Physics Research A, 718, 211, 2013.

[24] R.C.Jones, The General Theory of Bolometer Performance, Journal of the Optical Society of America, 43, 1. 1953.

[25] J.C.Mather, Bolometer Noise: non-equilibrium theory, Applied Optics, 21, 1125, 1982.

[26] J.C.Mather, Bolometer: ultimate sensitivity, optimization, and amplifier coupling, Applied Optics, 23, 584, 1984.

[27] M.Galeazzi, D.McCammon, Microcalorimeter and bolometer model, Journal of Applied Physics, 93, 4856, 2003.

[28] C.Arnaboldi, G.Boella, E.Panzeri, G. Pessina, JFET transistors for low noise applications at low frequency, IEEE Transaction on Nuclear Science, 51, 2975, 2004.

[29] D.Mccammon, M.Juda, J.Zhang, S.S.Holt, R.L.Kelley, S.H.Moseley, A.E.Szymkowiak, Thermal Detectors for High Resolution Spectroscopy, Japanese Journal of Applied Physics, 26, Supplement 26-3, 2084, 1987.

[30] S.Gaertner, A.Beno, J.M.Lamarre, M.Giard, J.L.Bret, J.P.Chabaud, F.X.Desert, J.P.Faure, G.Jegoudez, J.Lande, J.Leblanc, J.P.Lepeltier, J.Narbonne, M.Piat, R.Pons, G.Serra, G.Simiand, A new readout system for bolometers with improved low frequency stability, Astronomy & Astrophysics, 126, 151, 1997.

[31] M.E.Danowski, S.N.T.Heine, E.Figueroa-Feliciano, D.Goldfinger, P.Wikus, D.McCammon, P.Oakley, Vibration Isolation Design for the Micro-X Rocket Payload, Journal Low Temp Physics 184:597–603, 2016.

[32] A.Alessandrello, C.Brofferio, C.Bucci, E.Coccia, O.Cremonesi, E.Fiorini, A.Giuliani, A.Nucciotti, M.Pavan, G.Pessina, S.Pirro, E.Previtali, M.Vanzini, L.Zanotti, Vibrational and thermal noise reduction for cryogenic detectors, Nuclear Instruments and Methods in Physics Research, 444A, 331, 2000.

[33] P.Gorla, C.Bucci, S.Pirro, Complete elimination of 1K Pot vibrations in dilution refrigerators, Nuclear Instruments and Methods in Physics Research, A 520, 641, 2004.

[34] C.Arnaboldi, C.Bucci, O.Cremonesi, A.Fascilla, A.Nucciotti, M.Pavan, G.Pessina, S.Pirro, E.Previtali, M.Sisti, Low frequency noise characterization of very large value resistors, IEEE Transactioon Nuclear Science, 49, 1808, 2002.

[35] C.Arnaboldi, G.Pessina, The design of the input stage for the very front-end of the CUORE experiment, Journal of Low Temperature Physics, 151, 964, 2008.

[36] A.Alessandrello, C.Brofferio, C.Bucci, D.V.Camin, O.Cremonesi, A.Giuliani, A.Nucciotti, M.Pavan, G.Pessina, E.Previtali, A low dc drift read-out system for a large mass bolometric detector, IEEE Transaction in Nuclear Science, 44, 416, 1997.

[37] C.Arnaboldi, C.Bucci, J.W.Campbell, A.Nucciotti, M.Pavan, G.Pessina, S.Pirro, E.Previtali, C.Rosenfeld, M.Sisti, The programmable front-end system for cuoricino, an array of large-mass bolometers, IEEE Transactioon Nuclear Science, 49, 2440, 2002.

[38] C.Arnaboldi, X.Liu, G.Pessina, The preamplifier for CUORE, an array of large mass bolometers, 2009 IEEE Nuclear Science Symposium Conference Record N13-42.

[39] Sergio Franco, Design with operational amplifiers and analog integrated circuits, McGraw-Hill Education, 2002, from pag. 63.

[40] C.Arnaboldi, M.Cariello, S.DiDomizio, A.Giachero, G.Pessina, A programmable multichannel anti-aliasing filter for the CUORE experiment, Nuclear Instruments and Methods in Physics Research A, 617, 327, 2010.